\documentclass[a4paper,aps,showpacs,twocolumn]{revtex4}
\usepackage[pdftex]{graphicx}
\usepackage{amssymb}
\usepackage{natbib}

\begin{document}

\title{Directional point-contact
Andreev-reflection spectroscopy of Fe-based superconductors: Fermi
surface topology, gap symmetry,  and electron-boson interaction}
\author{D. Daghero, M. Tortello, G.A. Ummarino and R.S. Gonnelli}
\affiliation{Dipartimento di Fisica, Politecnico di Torino, 10129
Torino, Italy}

\begin{abstract}
Point-contact Andreev reflection spectroscopy (PCAR) has proven to
be one of the most powerful tools in the investigation of
superconductors, where it provides information on the order
parameter (OP), a fundamental property of the superconducting state.
In the past 20 years, successive improvements of the models used to
analyze the spectra have continuously extended its capabilities,
making it suited to study new superconductors with ``exotic''
properties such as anisotropic, nodal and multiple OPs. In Fe-based
superconductors, the complex compound- and doping-dependent Fermi
surface and the predicted sensitivity of the OP to fine structural
details present unprecedent challenges for this technique.
Nevertheless, we show here that PCAR measurements in Fe-based
superconductors carried out so far have already greatly contributed
to our understanding of these materials, despite some apparent
inconsistencies that can be overcome if a homogeneous treatment of
the data is used. We also demonstrate that, if properly extended
theoretical models for Andreev reflection are used, directional PCAR
spectroscopy can provide detailed information not only on the
amplitude and symmetry of the OPs, but also on the nature of the
pairing boson, and even give some hints about the shape of the Fermi
surface.
\end{abstract}
 \maketitle

\section{Introduction}
The order parameter (OP in the following) is the most important
quantity in a superconductor, since it determines how the charge
carriers couple to form Cooper pairs. In the standard BCS theory for
superconductivity \cite{bcs57}, the pairing potential is always
attractive, does not depend on energy, is isotropic in space and is
due to the interaction between electrons and phonons (in other
words, phonons are the ``mediating boson''). The amplitude of the
OP, $\Delta$, is also the ``energy gap'' in the superconductor, i.e.
the distance, in energy, between the condensate of Cooper pairs and
the first single-particle excitation. A dependence on energy of the
OP must be added, as in the so-called Eliashberg theory
\cite{eliashberg60}, to properly describe those superconductors,
like Pb, where the interaction between electrons and phonons is
particularly strong. In the past decades, many superconductors have
been discovered in which the OP is anisotropic (i.e. its amplitude
 shows some angular dependence) or can even
change sign to become \emph{negative} in some directions (as in the
so-called $d$-wave symmetry, common in copper-oxide
superconductors); in some cases, like the well-known MgB$_2$,
multiple OPs exist, related to different sets of bands that form
separated ``sheets'' of the Fermi surface.

The recently discovered Fe-based superconductors present most of
these unusual properties at the same time. The coupling of electrons
with the mediating boson (whatever its nature is) seems to be
intermediate to strong; the Fermi surface (FS) is made up of
separated sheets, which are mostly 2D-like; there are probably
multiple OPs that, in most cases, are expected to be isotropic but
with opposite sign on hole-like and electron-like FS sheets
\cite{mazin08}. In some conditions, OPs that display a sign change
on the same sheet of the FS are predicted to become energetically
favorable \cite{kuroki09,suzuki11}.

As usual when a new class of superconductors is discovered, the
experimental study of the OP in Fe-based superconductors has
immediately attracted the attention and expectations of the
scientific community. Determining the number, the amplitude and the
symmetry of the OPs is a necessary step to assess the coupling
mechanism and to understand the details of \emph{how}
superconductivity occurs in these materials. Point-contact
spectroscopy is one of the techniques that have been more useful in
the investigation of the OP in these materials. Despite its apparent
simplicity (it just consists in measuring the differential
conductance of a very small, in principle point-like contact between
a normal metal and the superconductor under study) it is a powerful
and versatile, yet inexpensive, research tool generally based on a
combination of quantum effects: the \emph{tunnel effect} (dominant
in high-resistance contacts) and the \emph{Andreev reflection}
(dominant in small-resistance contacts). Here we will concentrate on
the latter. To explain in a simple way what Andreev reflection is,
let's imagine an incident electron with total energy $E$ to travel
 towards the interface in the $N$ (normal) side of a ideal, barrierless $S-N$ junction
($S=$ superconductor). If it finds vacant electronic states at the
same energy on the $S$ side, the electron will simply propagate in
$S$. If instead its energy is smaller than the energy gap $\Delta$
in $S$, it cannot propagate in $S$, so it forms a Cooper pair with
another electron excited from the Fermi sea in $S$. This leaves a
hole in $S$ that crosses the interface and travels backwards in $N$
(see \cite{BTK,kashiwaya96,deutscher05} for more accurate
descriptions). From the point of view of $N$, things happen as if
the electron was reflected as a hole. This ``reflection'' causes a
doubling of the conductance with respect to a $N-N$ junction in
which Andreev reflection cannot take place. This doubling occurs
only as long as the energy of the injected electron (equal to $eV$
if $V$ is the voltage difference between $N$ and $S$) is smaller or
equal to $\Delta$. Thus, a measurement of the junction conductance
as a function of $V$ allows determining the gap, i.e. the amplitude
of the OP. In unconventional superconductors, the shape of the
conductance curves can provide information not only on the amplitude
of the OP, but also on its angular dependence (i.e. symmetry in
reciprocal space) and, in particular conditions, on the shape of the
Fermi surface and on the spectral function of the boson that
mediates the superconducting pairing, as we will discuss in section
\ref{subsect:FS}.

Point-contact Andreev-reflection (PCAR) spectroscopy measurements in
Fe-based superconductors have been carried out by various groups.
Early measurements suffered from the unsatisfactory quality of the
first samples and gave indeed contradicting and sometimes
unreproducible results. This might have given the impression that
PCAR does provide reliable results in these compounds. We will show
in section \ref{sect:PCAR} that, instead, when PCAR experiments in
good-quality samples are carried out with all the necessary cautions
by different groups and even in samples of different forms (single
crystals, polycrystals, films) they do provide surprisingly
consistent results. The sometimes contradicting conclusions arise
from the interpretation of the data rather than from the data
themselves. We will thus show that, in the most studied 1111 and 122
compounds, the available PCAR results agree rather well on the
amplitude of the OPs, on their temperature dependence and on their
symmetry (the $s\pm$ one, with isotropic OPs with sign reversal
between holelike and electronlike FS sheets being the most likely,
at least at optimal doping). In the same section we will also
present and discuss recent PCAR results in the 122 and 22426
families that seem to indicate a nodal, or strongly anisotropic OP.
If confirmed, these findings can serve as a test for the theories
that predict the occurrence of anisotropic and nodal symmetries in
suitable conditions, related to fine details of the lattice
structure.

The problem of extracting the electron-boson spectrum from PCAR
measurements in these materials will be dealt with in section
\ref{sect:EBIinFebased}. We will show that, with the combined use of
Eliashberg theory and of the models for PCAR conductance, it is
possible to extract the characteristic energy of the mediating boson
and also to pose some constraints on the shape of the relevant
spectrum. The strong electron-boson coupling also explains some
anomalies in the PCAR conductance curves, namely the excess
conductance at high energy, that greatly complicates the
normalization of the conductance curves and sometimes prevents a
good fit of the curves with models based on constant, BCS-like OPs.

Finally, in section \ref{sect:puzzle}, we will show that the values
of the gap ratios $2\Delta_i/k_B T_c$ collected from PCAR
measurements in various Fe-based superconductors seem to follow a
universal trend as a function of the critical temperature ($T_c$) of
the samples. Both $2\Delta_1/k_B T_c$ and $2\Delta_2/k_B T_c$ remain
approximately constant down to $T_c \simeq 30$ K, then tend to
\emph{increase} in compounds with lower $T_c$ -- even though further
measurements in the low-$T_c$ region would be necessary to confirm
this trend. We will propose a possible explanation of this behavior
that essentially relies on the fact that the mediating boson is an
electronic excitation (for example, spin fluctuations) and thus is
subject to a ``feedback effect'' of the condensate on the
electron-boson spectral function.

\section{Multiband 3D models for directional Andreev-reflection
spectroscopy}\label{sect:Models}
\subsection{The ``standard'' 1D and 2D BTK models}\label{subsect:BTK}

We already presented in a recent review \cite{daghero10} the
explicit expressions for the normalized conductance of a point
contact between a normal metal ($N$) and a superconductor ($S$) in
the case of a 2D geometry of the Fermi surfaces of the latter. Let
us briefly summarize these equations here, since in subsection
\ref{subsect:FS} we will use them as the starting point to extend
the Blonder-Tinkham-Klapwijk (BTK) model to particular 3D shapes of
the FS and to particular directions of current injection. In the
following we will call ``2D BTK model'' the generalization of the
original, ``1D'' BTK theory \cite{BTK}, introduced in 1996 by Y.
Tanaka and S. Kashiwaya \cite{kashiwaya96}. In their article they
first wrote the normalized conductance
$G(E)=(dI/dV)_{\mathrm{NS}}/(dI/dV)_{\mathrm{NN}}$ at $T=0$ (where
$I_{\mathrm{NS}}$ and $I_{\mathrm{NN}}$ are the currents flowing
through the interface when the material under study is in the
superconducting and in the normal state, respectively) as a function
of the two quantities
$N_{\mathrm{q}}(E)=E/\sqrt{E^{2}-\Delta{}^{2}}$ and
$N_{\mathrm{p}}(E)=\Delta/\sqrt{E^{2}-\Delta^{2}}$, whose real parts
are the BCS quasiparticle and pair density of states, respectively.

Here we will immediately generalize these expressions to the
strong-coupling regime, where the order parameter $\Delta$ is a
complex function of energy resulting from the solution of the
Eliashberg equations, and to the presence of inelastic quasiparticle
scattering processes at the N/S interface. In the latter case it has
been shown \cite{plecenik94} that the resulting (and dominant)
extrinsic reduction of the quasiparticle lifetime can be properly
taken into account by including in the model an imaginary part of
the energy described by the broadening parameter $\Gamma$, i.e.
$E\rightarrow E+i\Gamma.$ As a consequence, from now on:
$N_{\mathrm{q}}(E,\Gamma)=(E+i\Gamma)/\sqrt{(E+i\Gamma)^{2}-\Delta(E)^{2}}$
and
$N_{\mathrm{p}}(E,\Gamma)=\Delta(E)/\sqrt{(E+i\Gamma)^{2}-\Delta(E)^{2}}$.
Note that we have allowed $\Delta$ to depend on energy for the sake
of generality, even though we will use this dependence explicitly
only in section \ref{subsect:EBI}.

The main steps in the calculation of the normalized conductance
$G(E)$ consist in the determination of the barrier transparency when
the material under study is in the normal state
($\tau_{\mathrm{N}}$) and in the superconducting state
($\sigma_{\mathrm{S}}(E)$). These quantities are actually
transmission probability distributions and are obtained by solving
the Bogoliubov-De Gennes equations with the only condition that the
component of the wavevector $\mathbf{k}$ parallel to the interface
is conserved in all processes, for any direction of the incoming
electron from the N side. In the pure 2D case here discussed, where
we suppose the interface to coincide with a plane parallel to the
\emph{z} axis, the incoming electrons lie in the \emph{xy} plane and
have a wave vector $\mathbf{k}_{\mathrm{F,N}}$ which forms an angle
$\theta_{\mathrm{N}}$ with respect to the unitary vector
\emph{$\mathbf{n}$} normal to the interface. In principle, if the
Fermi velocity is different in the N and S side, the charge carriers
experience a ``refraction'' when crossing the interface (i.e.  the
wavevector of the quasiparticles in the S side of the junction
$\mathbf{k}_{\mathrm{F,S}}$ could be different from
$\mathbf{k}_{\mathrm{F,N}}$) so that the direction of propagation of
quasiparticles in the S side could form an angle
$\theta_{\mathrm{S}}$ different from $\theta_{\mathrm{N}}$ with
respect to the normal to the interface. In these conditions,
electrons incident on the interface with suitable angles can
experience a normal reflection even in absence of a barrier
potential $U_{0}$ located at the interface, and the normal
transmission coefficient $\tau_{\mathrm{N}}$ is no longer
identically 1. In the most general case when $U_{0}$ is present, the
normal transparency of the barrier is:
\begin{equation}
\tau_{\mathrm{N}}=\frac{4(\mathbf{k}_{\mathrm{F,N}}\cdot\mathbf{n})%
(\mathbf{k}_{\mathrm{F,S}}\cdot\mathbf{n})}{(\mathbf{k}_{\mathrm{F,N}}%
\cdot\mathbf{n}+\mathbf{k}_{\mathrm{F,S}}\cdot\mathbf{n})^{2}+4Z^{2}k_{\mathrm{F,N}}^{2}}
\label{eq_tauN_general}
\end{equation}
where $Z=U_{0}/\hbar v_{\mathrm{F}}$ is the dimensionless parameter
which accounts for the potential barrier.

From now on we will restrict the analysis to the case
$\mathbf{k}_{\mathrm{F,N}}$=$\mathbf{k}_{\mathrm{F,S}}=\mathbf{\mathbf{k}_{\mathrm{F}}}$
from which it follows
$\theta_{\mathrm{N}}$=$\theta_{\mathrm{S}}$=$\theta$. As a matter of
fact, while properly taking  into account the Fermi velocity
mismatch at the interface in the 2D case is a non-straightforward
but feasible task (see \cite{daghero10}), it becomes an extremely
complex problem in the full 3D case we will discuss in subsection
\ref{subsect:FS}, when the true shape of the FS will be considered.
Under the previous restrictive hypothesis the normal transparency
becomes:
\begin{equation}
\tau_{\mathrm{N}}(\theta)=\frac{\cos^{2}\theta}{\cos^{2}\theta+Z^{2}}.\label{eq:tauN}
\end{equation}

It is clear that, for $Z=0$, the normal transmission probability is
identically 1 for any direction of the incoming electron, i.e. for
$-\pi/2<\theta<\pi/2$. When $Z\neq 0$, instead, the barrier
transparency depends on the direction of the incoming electron in
the N side; for $Z=10$ (tunneling regime) the transmission
probability is always small and highly directional. It can be shown
that the relative transparency of the barrier in the superconducting
state is given by \cite{kashiwaya96}:
\begin{equation}
\sigma_{\mathrm{S}}(E,\theta)=\frac{1+\tau_{\mathrm{N}}(\theta)|\gamma(E)|^{2}+(\tau_{\mathrm{N}}(\theta)-1)|\gamma(E)^{2}|^{2}}{|1+\left[\tau_{\mathrm{N}}(\theta)-1\right]\gamma(E)^{2}|^{2}}
\label{eq:sigma(E_theta)}
\end{equation}
where $\gamma(E)=\left[N_{\mathrm{q}}(E)-1\right]/N_{\mathrm{p}}(E)$
and finally the normalized conductance at $T=0$ is:
\begin{equation}
G_{2\mathrm{D}}(E)=\frac{\intop_{-\frac{\pi}{2}}^{+\frac{\pi}{2}}\sigma_{\mathrm{S}}(E,\theta)\tau_{\mathrm{N}}(\theta)\cos\theta
d\theta}{\intop_{-\frac{\pi}{2}}^{+\frac{\pi}{2}}\tau_{\mathrm{N}}(\theta)\cos\theta
d\theta}. \label{eq:G2D}
\end{equation}

Both at numerator and denominator of equation \ref{eq:G2D} the
integration extremes are $-\pi/2$ and $\pi/2$ since the current
injection occurs in the whole positive half-space for which
$0\leq\cos\theta\leq1$. The standard BTK theory \cite{BTK} is the 1D
limit of this model, in which all the electrons from the N side are
injected in the same direction perpendicular to the interface.
Obviously, the normal-state transmission probability in this case is
simply obtained by putting $\theta=0$ in equation \ref{eq:tauN},
i.e. $\tau_{\mathrm{N}}=1/(1+Z^{2})$. By doing the same also in
equations \ref{eq:sigma(E_theta)} and \ref{eq:G2D} the expression
for the 1D BTK conductance
$G_{1\mathrm{D}}(E)=\sigma_{\mathrm{S}}(E,0)$ is thus recovered.
Instead, the 2D BTK model itself is the limit of the more general 3D
model (see section \ref{subsect:FS}) if the gap is isotropic and the
FS is spherical (i.e. the system has a rotational symmetry around
the axis normal to the interface).
\begin{figure}[t]
  \includegraphics[width=\columnwidth]{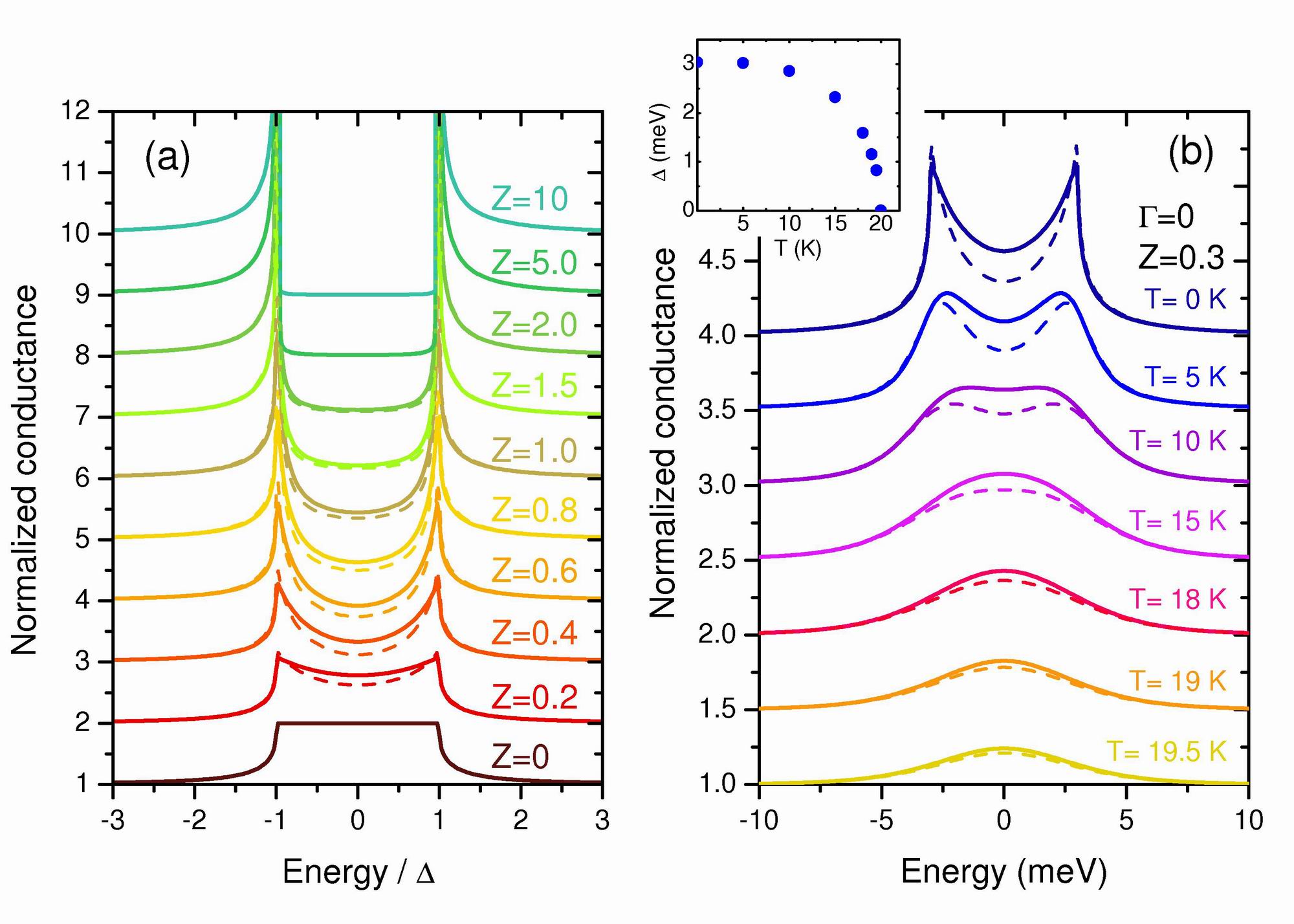}\\
  \caption{(color online): (a) Theoretical PCAR normalized conductance
of a single-band \emph{s}-wave superconductor as function of reduced
energy $E/\Delta$ at $T=0$ and $\Gamma=0$ for different $Z$ values
from the pure Andreev-reflection regime ($Z=0$) to the tunneling one
($Z=10$). The solid curves are calculated by using the 1D BTK model,
while the dashed ones by using the 2D BTK one (see text). (b)
Temperature dependency of the theoretical PCAR normalized
conductance of panel (a) for $\Delta=3$ meV, $\Gamma=0$ and $Z=0.3$.
Also in this case solid curves represent the results of 1D BTK model
and dashed ones the results of 2D BTK one. In the inset the BCS
temperature dependency of the gap is shown. The curves in (a) and
(b) are vertically offset for clarity.}\label{fig:1D_2D}
\end{figure}

In Figure \ref{fig:1D_2D}(a) a comparison of the normalized
conductances $G_{1\mathrm{D}}(E)$ (solid lines) and $G_{2D}(E)$
(dashed lines) is shown at $T=0$ and $\Gamma=0$ as a function of
the normalized energy $E/\Delta$ for different values of the
barrier parameter $Z$. Looking at the form of
$\tau_{\mathrm{N}}(\theta)$ (equation \ref{eq:tauN}), one expects
the 1D BTK normalized conductance to coincide with the 2D one in
two opposite conditions: when $Z=0$ (i.e. when
$\tau_{\mathrm{N}}=1$ and
$\sigma_{\mathrm{S}}(E)=1+|\gamma(E)|^{2}$) and when
$Z\rightarrow\infty$ (i.e. when $\tau_{\mathrm{N}}\rightarrow0$
and
$\sigma_{\mathrm{S}}(E)\rightarrow(1-|\gamma(E)^{2}|^{2})/|1-\gamma(E)^{2}|^{2}$).
This is clearly shown in the curves at $Z=10$ and $Z=0$. The
maximum deviations between the predictions of the two models occur
for $Z \simeq 0.3-0.5$ which, incidentally, is the range of values
commonly observed in experiments in Fe-based superconductors. It
is very easy to demonstrate that, in practice, each $G_{2D}(E)$
curve almost coincides with the $G_{1D}(E)$ calculated for a
properly enhanced $Z$ value and, thus, the 1D BTK model can be
efficiently used instead of the 2D one provided that the gap is
isotropic, the FS spherical and one is not interested in the
precise determination of the $Z$ values.

All the equations seen up to now are calculated at $T=0$. The effect
of a finite temperature $T$ can be properly taken into account by
simply convolving the normalized conductance with the Fermi function
at that temperature, as shown in detail in \cite{daghero10} and by
including in the calculations the proper temperature dependence of
the gap. Figure \ref{fig:1D_2D}(b) shows the temperature dependence
of the normalized conductances $G_{1\mathrm{D}}(E)$ (solid lines)
and $G_{2D}(E)$ (dashed lines) for $\Delta=3.0$ meV, $\Gamma=0$,
$Z=0.3$ and $T_{\mathrm{c}}=20$ K. The BCS temperature dependence of
the gap $\Delta(T)$ used in the calculations is shown in the inset.
This kind of curves is what is expected in the best PCAR experiments
($\Gamma=0$) on a single-band isotropic BCS superconductor.

\subsection{Symmetry of the order parameter}\label{subsect:symmetry}
In order to understand the profound changes occurring in the
normalized conductance of a point contact when the symmetry of the
order parameter is not \emph{s} wave, it is necessary to briefly
describe in greater detail the physical processes that take place at
the NS interface. When an electron coming from the N side reaches
the potential barrier at the interface, four different processes can
occur depending on the electron energy and on the height of the
barrier: A) the electron is reflected as a hole in N (Andreev
reflection); B) the electron is specularly reflected as an electron
in N (normal reflection); C) the electron is transmitted in S as an
electron-like quasiparticle, ELQ (normal transmission); D) the
electron is transmitted in S as a hole-like quasiparticle, HLQ
(anomalous transmission). Due to the charge conservation this last
process results in two electrons ``reflected'' back in N for every
incoming electron. The probabilities of these processes strongly
depend on the energy of the incoming electron and on the $Z$
parameter. In particular, processes C and D are forbidden for
$E<\Delta$ (where there can be no transmitted quasiparticles),
process B is present at any energy when $Z>0$ and process A is also
present at any energy (but is very small for $E>\Delta$) only if $Z$
values are not too large (in practice $Z<10$). Equation \ref{eq:G2D}
takes properly into account all these probabilities, but only in the
case of an isotropic order parameter, i.e. in \emph{s}-wave
symmetry. When the OP is anisotropic, i.e. its amplitude and
possibly its sign change as a function of the direction in
reciprocal space, the situation is more complex. As a matter of
fact, the wavevector of the ELQ $\mathbf{k}_{\mathrm{F}}^{+}$ and
that of the HLQ $\mathbf{-k}_{\mathrm{F}}^{-}$ have now different
direction, yet still having the same component parallel to the
interface. In other words, the two vectors in the S side form the
angles $\theta$ and $-\theta$ with respect to the unitary vector
$\mathbf{n}$, respectively \cite{kashiwaya96}. In this situation and
depending on the orientation of the anisotropic OP with respect to
$\mathbf{n}$ the ELQs and the HLQs can feel different OPs.

Let us extend the equations for the 2D BTK model to this case. Let
us assume that the system has a translational invariance along
$\mathbf{k}_{\mathrm{z}}$, i.e. the FS is a cylinder, the OP has a
$\mathbf{k}$ dependence only in the $k_{\mathrm{x}}k_{\mathrm{y}}$
plane (as it happens in cuprates) and the current injection occurs
in the same plane ($ab$-plane contact). We can now introduce the
angle of misalignment $\alpha$ between the crystallographic \emph{a}
axis in S and the normal to the interface $\mathbf{n}$.
%
As a consequence of gap anisotropy and of this misalignment, ELQs
feel the order parameter
$\Delta(E,\theta)_{+}=\Delta(E,\theta-\alpha)$, while HLQs feel
$\Delta(E,\theta)_{-}=\Delta(E,-\theta-\alpha)$. In this case the
relative transparency of the barrier in the superconducting state
becomes \cite{kashiwaya96}:
\begin{eqnarray}
& &\sigma_{\mathrm{S}}(E,\theta)= \label{eq:sigma_anis}\\
& &
\frac{1+\tau_{\mathrm{N}}(\theta)|\gamma_{+}(E,\theta)|^{2}+(\tau_{\mathrm{N}}(\theta)-1)|\gamma_{+}(E,\theta)\gamma_{-}(E,\theta)|^{2}}{|1+(\tau_{\mathrm{N}}(\theta)-1)\gamma_{+}(E,\theta)\gamma_{-}(E,\theta)\exp(i\varphi_{d})|^{2}}\nonumber
\end{eqnarray}
where
\begin{equation}
\gamma_{\pm}(E,\theta)=\frac{(E+i\Gamma)-\sqrt{(E+i\Gamma)^{2}-|\Delta(E,\theta)_{\pm}|^{2}}}{|\Delta(E,\theta)_{\pm}|}
\end{equation}
and
\begin{equation}
\varphi_{d}(\theta)=-i\ln\left[\frac{\Delta(E,\theta)_{+}/\left|\Delta(E,\theta)_{+}\right|}{\Delta(E,\theta)_{-}/\left|\Delta(E,\theta)_{-}\right|}\right]
\end{equation}
is the phase difference (function of $\theta$) seen by the HLQs with
respect to the ELQs. Of course, $\tau_{\mathrm{N}}(\theta)$ remains
the same (indeed it only describes processes B and C in a N-N
junction) and thus, simply by introducing the new expression for
$\sigma_{\mathrm{S}}(E,\theta)$ (equation \ref{eq:sigma_anis}) into
equation \ref{eq:G2D} and performing the proper integration in
$\theta$ one can obtain the normalized conductance at $T=0$ for any
symmetry of the OP in the $k_{\mathrm{x}}k_{\mathrm{y}}$ plane. This
symmetry is of course expressed by the specific $\theta$ dependence
of the functions $\Delta(E,\theta)_{\pm}$.

\begin{figure*}[t]
\includegraphics[width=0.7\textwidth]{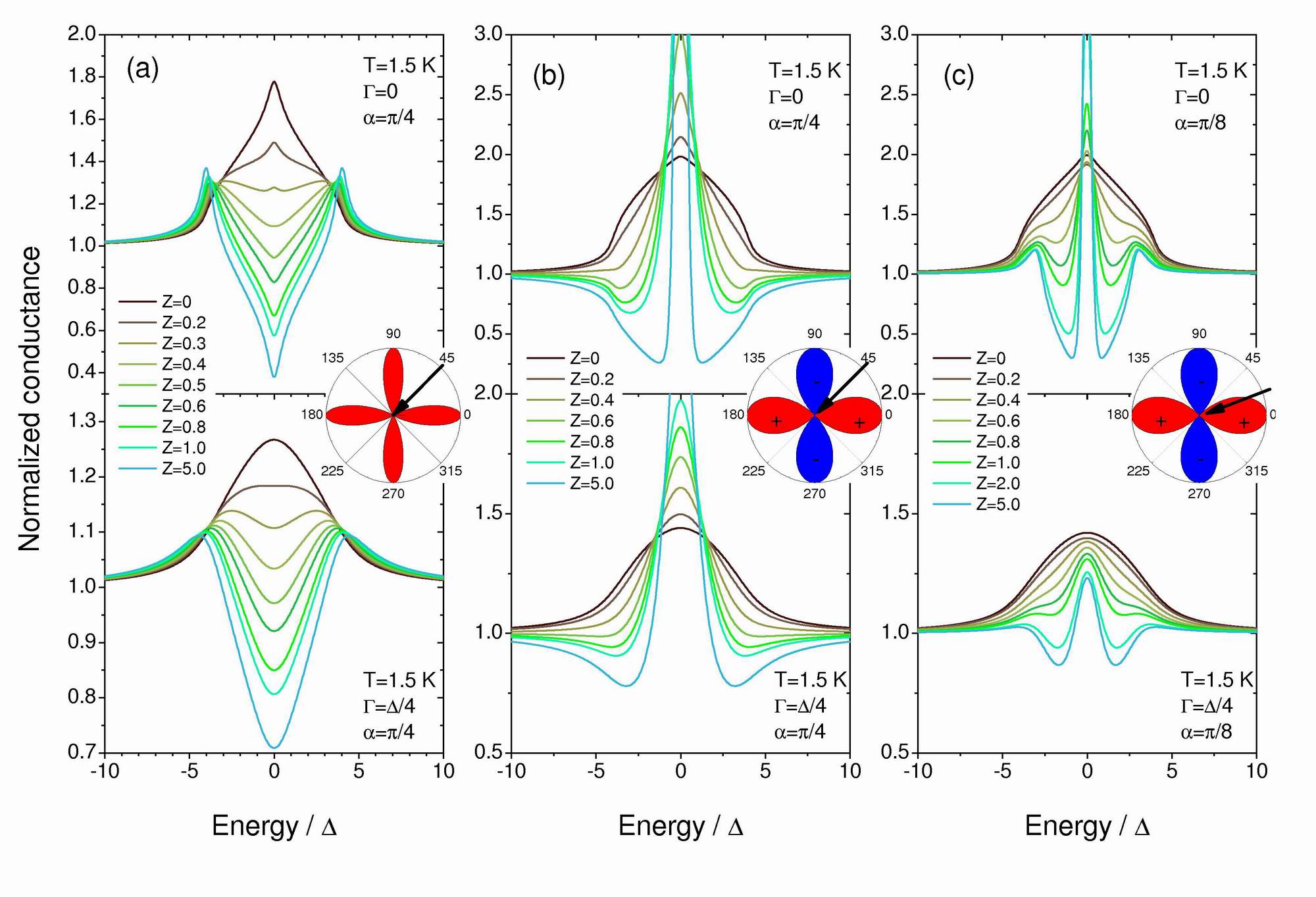}\\
\caption{(color online): (a) Reduced energy dependence of the
normalized conductance of a single-band superconductor with full
anisotropic \emph{s} symmetry of the OP for different $Z$ values at
$T=1.5$ K,$\alpha=\pi/4$ and $\Gamma=0$ (upper panel) and at $T=1.5$
K,$\alpha=\pi/4$ and $\Gamma=\Delta/4$ (lower panel); (b) The same
as in panel (a) but for a single-band superconductor with
$d_{\mathrm{x^{2}-y^{2}}}$ symmetry of the OP; (c) The same as in
panel (b) but for $\alpha=\pi/8$. The insets in all the panels
represent the symmetry of the pair potential (red and blue colors
mean positive and negative values, respectively) while the black
arrows indicate the direction of current injection.}\label{fig:2}
\end{figure*}

In the present review we are interested in the symmetries that have
been considered plausible for Fe-based superconductor. The $s\pm$
symmetry, with isotropic OPs of different sign on the hole-like and
on the electron-like FS sheets \cite{mazin08}, is the most likely
for the majority of the systems. If interband interference effects
\cite{golubov09} are negligible, this phase change is not detectable
by PCAR and gives the same spectra as a multiband $s$-wave symmetry.
In suitable conditions, that seem to be related to fine structural
parameters \cite{kuroki09,suzuki11}, anisotropic or nodal symmetries
are expected to become more energetically favorable. As a
consequence, apart from the \emph{s}-wave, we will discuss here the
cases of \emph{fully anisotropic s-wave} symmetry (with zeros in the
gap) and $d_{x^{2}-y^{2}}$\emph{-wave }one. The two expressions for
the OP are:
$\Delta_{\mathrm{an}}(E,\theta)_{\pm}=\Delta(E)\cos^{4}\left[2(\pm\theta-\alpha)\right]
\label{eq:s_anis}$
and
$\Delta_{\mathrm{d}}(E,\theta)_{\pm}=\Delta(E)\cos2(\pm\theta-\alpha),
\label{eq:gap_d}$
respectively \cite{vanharlingen,kashiwaya96}. In the first case the pair potential is zero in four
directions but never changes sign, while in the second case the sign
change typical of \emph{d}-wave symmetry is present, as shown in the
insets of figure \ref{fig:2}.

Figure \ref{fig:2}(a) reports the theoretical 2D AR normalized
conductance curves obtained from the integration of equation
\ref{eq:G2D} for \emph{fully anisotropic s}-wave symmetry, at finite
temperature ($T=1.5$ K)  and different $Z$ values. Upper and lower
panels refer to the cases $\Gamma=0$ and $\Gamma=\Delta /4$,
respectively. Figure \ref{fig:2}(b) and (c) report the corresponding
conductance curves obtained for a $d$-wave symmetry, when
$\alpha=\pi/4$ and $\alpha=\pi/8$, respectively. Without entering
too much into details, the main message coming from these
calculations is the following. For current injection in the
\emph{ab} plane, and when $Z$ is sufficiently small, a maximum of
the conductance at zero bias appears both in \emph{anisotropic s}-
and \emph{d}-wave symmetries (top panels). These maxima however have
a somehow different physical origin in the two symmetries. In
\emph{fully anisotropic s} symmetry, they come only from the angular
integration over OPs that can assume any value between the maximum
$\Delta(E)$ and 0. In \emph{d}-wave symmetry, the zero-bias maximum
has the same origin only for $Z=0$, while for finite $Z$ values it
considerably grows (becoming much higher than 2 in the tunneling
regime) because of an additional effect, i.e. the constructive
interference of ELQs and HLQs that feel a phase difference of the
pair potential. This interference effect (described by the
$\exp(i\varphi_{d})$ term in the denominator of equation
\ref{eq:sigma_anis}) is the only responsible for the zero-bias
conductance peak (ZBCP) present in the \emph{d}-wave case for
$\alpha>0$ and large $Z$, a peak that is totally absent in the
\emph{anisotropic s} case.

In principle, the different shape of the conductance curves for the
two symmetries at low $Z$ shown in the upper panels of figure
\ref{fig:2} would allow one to identify the true OP symmetry. In
practice, however, even a relatively small amount of broadening,
i.e. $\Gamma= \Delta/4$, is sufficient to prevent the correct
determination of the OP symmetry, as shown in the lower panels of
figure \ref{fig:2}, unless $Z$ is sufficiently large. At low $Z$,
the ZBCP typical of the $d$-wave symmetry, smoothed by $\Gamma$, can
be confused with the zero-bias maximum of the \emph{anisotropic
s}-wave one. Thus, in  the presence of a zero-bias maximum in the
experimental curves of \emph{ab}-plane point contacts in AR regime
(i.e at low $Z$) what we can only say is that the OP is zero along
some directions. Only \emph{ab}-plane measurements in the
\emph{tunneling} regime would allow us to distinguish the two
situations but, apparently, they are not so easy to obtain in
Fe-based compounds with point-contact or STM techniques.

\subsection{Shape of the Fermi surface and directionality}\label{subsect:FS}
In this subsection we will extend the 2D BTK theory to arbitrary
shapes of the FS and will show what is expected when the current is
injected either in the $ab$ plane or along the $c$ axis. Under a few
restrictive hypotheses (still present in this approach), the final
result will be the full 3D generalization of the BTK model to any
anisotropic feature both of the FS and of the pair potential
symmetry.

The first obvious generalization, essential for the interpretation
of the results in Fe-based superconductors, is to consider the
multiband nature of superconductivity in these compounds and the
multiple sheets of the FS arising from the crossings of the Fermi
energy by the different bands. This problem gained for the first
time importance and popularity ten years ago with the discovery of
$\mathrm{MgB_{2}}$, the first well-known multiband, multigap
superconductor \cite{gonnelli02c,brinkman02}. At the level of AR
data analysis and in order to have a limited number of fitting
parameters the standard approach is to consider a two-band, two-gap
system where the values of $\Delta_{\mathrm{i}}$, $Z_{\mathrm{i}}$
and $\Gamma_{\mathrm{i}}$ of every band ($\mathrm{i}=1,2)$ have to
be determined by the best fit of theoretical curves to the
experimental ones. The total normalized conductance is thus written
as: $G_{\mathrm{tot}}(E)=w_{1}G_{1}(E)+(1-w_{1})G_{2}(E)$ where
$G_{\mathrm{i}}(E)$ is the normalized conductance of the i-th band
(1D or 2D depending on the selected model) and $w_{1}$ is the
relative contribution of band 1 to the total conductance. The
drawback here is that, apart from a few lucky cases
\cite{brinkman02}, the weight factor $w_{1}$ is also a parameter of
the fit. We will see later that this is not the case in the full 3D
model where the weighting factors come automatically from the
complex interaction between the FS shape and the direction of
current injection.

A simple extension of the 2D BTK theory to the 3D case can be
realized under the hypothesis of a spherical FS in the S side of the
junction (the N side is always supposed to have a spherical FS) as
shown in \cite{yamashiro97}. The main point here is to express
$\tau_{\mathrm{N}}$, $\sigma_{\mathrm{S}}(E)$ and $\Delta(E)$ (if
the OP is anisotropic) as a function of both the azimuthal and
inclination angles $\theta$ and $\phi$, respectively. Of course the
specific expressions depend on the direction of current injection
and on the particular symmetry of the pair potential. As an example,
for a \emph{a}-axis contact (always with no mismatch of Fermi
velocity at the interface) it is rather easy to show that:
\[
\tau_{\mathrm{N}x}^{sph}(\theta,\phi)=\frac{\cos^{2}\theta\sin^{2}\phi}{\cos^{2}\theta\sin^{2}\phi+Z^{2}}
\]
and $\Delta(E,\theta,\phi)_{\pm}$ are the expressions of the pair
potential seen by the ELQs and HLQs, respectively, for the specific
symmetry considered. By introducing $\Delta(E,\theta,\phi)_{\pm}$
into the above formulas for $\gamma_{\pm}$ and $\varphi_{d}$
(equations 6 and 7), and then inserting
$\tau_{\mathrm{N}x}^{sph}(\theta,\phi)$ and the obtained
$\gamma_{\pm}(E,\theta,\phi)$ and $\varphi_{d}(\theta,\phi)$ into
equation \ref{eq:sigma_anis} it is finally possible to integrate on
the two angles arriving at the normalized conductance:
\begin{equation}
G_{3\mathrm{D}x}^{sph}(E)=\frac{\intop_{0}^{\pi}\intop_{-\frac{\pi}{2}}^{+\frac{\pi}{2}}\sigma_{\mathrm{S}x}^{sph}(E,\theta,\phi)\tau_{\mathrm{N}x}^{sph}(\theta,\phi)\cos\theta\sin^{2}\phi
d\theta
d\phi}{\intop_{0}^{\pi}\intop_{-\frac{\pi}{2}}^{+\frac{\pi}{2}}\tau_{\mathrm{N}x}^{sph}(\theta,\phi)\cos\theta\sin^{2}\phi
d\theta d\phi}.\label{eq:G3D}\end{equation}

The model discussed so far is still implicitly based on the
assumption of a spherical FS. The problem then arises of how to
express the \emph{c}-axis or \emph{a}-axis conductance in a
superconductor with a quasi-2D FS.  This is the case of cuprates or
of $\mathrm{Sr_2RuO_4}$. A possible solution, though not really
general, is to restrict the integration in $\phi$ to a narrow
angular range around $\pi/2$ as in \cite{yamashiro97}. In this way,
the spherical FS of the model is reduced to a almost-cylindrical
belt that locally approximates the true, cylindrical FS. In the
following, we will instead derive and present a more general
approach to a full 3D case, with a FS of arbitrary shape.

Just for simplicity we will still limit the calculations to the
contribution of only two bands (thus the band index $\mathrm{i}$
assumes only the values 1 and 2). As for the directionality of the
contact, we consider here current injections along the \emph{x} axis
(\emph{yz}-plane interface) and along the \emph{z} axis
(\emph{xy}-plane interface), being representative of the typical
experimental conditions in single crystals of the Fe-based
compounds, i.e. \emph{ab}-plane and \emph{c}-axis contacts,
respectively. From now on the subscript \emph{x} or \emph{z} will
denote the specific expressions to be used for the two contact
directions.

The particular shape of the i-th FS is described by the wave vector
$\mathbf{k}_{\mathrm{F},\,\mathrm{i}}(\theta,\phi)$, while the
unitary vector perpendicular to the FS at any point of the
reciprocal space is simply given by
$\mathbf{n}_{\mathrm{F},\,\mathrm{i}}(\theta,\phi)=\frac{\partial\mathbf{k}_{\mathrm{F},%
\,\mathrm{i}}(\theta,\phi)}{\partial\theta}\times\frac{\partial\mathbf{k}_{\mathrm{F},\,%
\mathrm{i}}(\theta,\phi)}{\partial\phi}$ whose calculation is
trivial once the shape of $i$-th FS has been chosen.

Under the additional (but reasonable) hypothesis that the points on
the FS are close to the points of maximum symmetry of the energy
bands (i.e. they are close to the top or the bottom of
parabolic-like bands) one can express the Fermi velocity at any
point as a function of the component of
$\mathbf{k}_{\mathrm{F},\,\mathrm{i}}$ perpendicular to the FS at
that point:

\begin{eqnarray}
\mathbf{v}_{\mathrm{F},\,\mathrm{i}}(\theta,\phi)&=&\frac{\hbar
k_{F,\,
i}^{\bot}(\theta,\phi)}{m^*}\mathbf{n}_{\mathrm{F},\,\mathrm{i}}(\theta,\phi)\nonumber \\
&= &\frac{\hbar\mathbf{k}_{F,\,
i}(\theta,\phi)\cdot\mathbf{n}_{\mathrm{F},\,\mathrm{i}}(\theta,\phi)}{m^*}\mathbf{n}_{\mathrm{F},\,\mathrm{i}}(\theta,\phi)
\nonumber
\end{eqnarray}
where $m^*$ is the effective mass of quasiparticles. Finally, the
components of the Fermi velocity along a given direction can be
simply obtained by projecting
$\mathbf{v}_{\mathrm{F},\,\mathrm{i}}(\theta,\phi)$ along that
direction, i.e.
$v_{\mathrm{F}x,\,\mathrm{i}}(\theta,\phi)=\mathbf{v}_{\mathrm{F},\,\mathrm{i}}(\theta,\phi)\cdot\mathbf{\hat{i}}$
and
$v_{\mathrm{F}z,\,\mathrm{i}}(\theta,\phi)=\mathbf{v}_{\mathrm{F},\,\mathrm{i}}(\theta,\phi)\cdot\mathbf{\hat{k}}$
where, of course, $\mathbf{\hat{i}}$ and $\mathbf{\hat{k}}$ are the
unitary vectors of the $x$ and $z$ axis, respectively. We have now
all the ingredients required to calculate the normalized
conductance.

By following the approach originally developed in
\cite{mazin99,brinkman02} and recently summarized in
\cite{daghero10}, and by neglecting possible interference effects
between bands that can lead to the formation of bound states at the
surface \cite{golubov09}, we obtain the normalized conductance for
current injection along the $x$ axis:

\begin{widetext}
\begin{equation}
G_{\mathrm{3D}x}(E)=\frac{\sum_{i}\intop_{\phi_{\mathrm{min}}}^{\pi-\phi_{\mathrm{min}}}\intop_{-\frac{\pi}{2}}^{+\frac{\pi}{2}}\sigma_{\mathrm{S}x,\,\mathrm{i}}(E,\theta,\phi)
\tau_{\mathrm{N}x,\,\mathrm{i}}(\theta,\phi)
\frac{v_{\mathrm{F}x,\,\mathrm{i}}(\theta,\phi)}{v_{\mathit{\mathrm{F},\,
i}}(\theta,\phi)}k_{F,\, \mathrm{i}}^2(\theta,\phi)\sin \phi d\theta
d\phi}{\sum_{i}\intop_{\phi_{\mathrm{min}}}^{\pi-\phi_{\mathrm{min}}}\intop_{-\frac{\pi}{2}}^{+\frac{\pi}{2}}\tau_{\mathrm{N}x,\,\mathrm{i}}(\theta,\phi)
\frac{v_{\mathrm{F}x,\,\mathrm{i}}(\theta,\phi)}{v_{\mathit{\mathrm{F},\,
i}}(\theta,\phi)} k_{F,\, \mathrm{i}}^2(\theta,\phi)\sin \phi
d\theta d\phi} \label{eq:G3Dx}
\end{equation}
\end{widetext}
where
\begin{equation}
\tau_{\mathrm{N}x,\,\mathrm{i}}(\theta,\phi)=\frac{4v_{\mathrm{F}x,\,\mathrm{i}}(\theta,\phi)v_{\mathrm{N}x}}{\left[v_{\mathrm{F}x,\,\mathrm{i}}(\theta,\phi)+v_{\mathrm{N}x}\right]^{2}+4Z_{x,\,\mathrm{i}}^{2}v_{\mathrm{N}}^{2}}
\label{eq:tau_3Dx}
\end{equation}
and $\sigma_{\mathrm{S}x,\,\mathrm{i}}(E,\theta,\phi)$ has the same
expression shown in equation \ref{eq:sigma_anis} but now contains
the band-associated functions
$\gamma_{\pm,\,\mathrm{i}}(E,\theta,\phi)$, obtained by substituting
the OP expressions $\Delta(E,\theta,\phi)_{\pm,\,\mathrm{i}}$ into
the $\gamma_{\pm}$ expressions (equation 6 of subsection
\ref{subsect:symmetry}), and the band-associated,
direction-dependent normal transparency
$\tau_{\mathrm{N}x,\,\mathrm{i}}(\theta,\phi)$.

Equation \ref{eq:tau_3Dx} directly comes from equation
\ref{eq_tauN_general} if the difference in the effective masses of
all the particles and quasiparticles involved in the process (in S
and N) is disregarded. The condition of no mismatch of Fermi
velocities across the interface translates here into
$v_{\mathrm{F},\,\mathrm{i}}(\theta^*,\phi^*)=v_{\mathrm{N}}$ where
$\theta^*$ and $\phi^*$ define a point on the FS (usually in the
$k_{x}k_{y}$ plane) where we suppose the two velocities to be equal.
By imposing this condition, we eliminate the need to know the term
$\hbar/m^*$ which multiplies all the elements at numerator and
denominator of equations \ref{eq:G3Dx} and \ref{eq:tau_3Dx}. In this
way, of course, we neglect the small deviation of the quasiparticles
in crossing the interface due to the modest mismatch of Fermi
velocities in N and S simply arising from the  geometry of the FS in
S. In equations \ref{eq:G3Dx} and \ref{eq:tau_3Dx}
$v_{\mathrm{N}x}=\mathbf{v}_{\mathrm{N}}\cdot\mathbf{\hat{i}}=v_{\mathrm{N}}\cos\theta\sin\phi$
and $Z_{x,\,\mathrm{i}}$ is the barrier height parameter for current
injection along the \emph{x} axis in the $i$-th band. Finally, the
limits of integration in $\phi$ ($\phi_{min}$ and $\pi-\phi_{min}$)
are fixed in such a way that the integration is restricted to the
first Brillouin zone.

If all the previous considerations hold, writing the expression of
the normalized conductance for current injection along the $z$ axis
is straightforward:
\begin{widetext}
\begin{equation}
G_{\mathrm{3D}z}(E)=\frac{\sum_{i}\intop_{\phi_{\mathrm{min}}}^{\pi/2}\intop_{0}^{2\pi}\sigma_{\mathrm{S}z,\,\mathrm{i}}(E,\theta,\phi)\tau_{\mathrm{N}z,\,\mathrm{i}}(\theta,\phi)
\frac{v_{\mathrm{F}z,\,\mathrm{i}}(\theta,\phi)}{v_{\mathit{\mathrm{F},\,
i}}(\theta,\phi)} k_{F,\, \mathrm{i}}^2(\theta,\phi)\sin \phi
d\theta
d\phi}{\sum_{i}\intop_{\phi_{\mathrm{min}}}^{\pi/2}\intop_{0}^{2\pi}\tau_{\mathrm{N}z,\,\mathrm{i}}(\theta,\phi)
\frac{v_{\mathrm{F}z,\,\mathrm{i}}(\theta,\phi)}{v_{\mathit{\mathrm{F},\,
i}}(\theta,\phi)} k_{F,\, \mathrm{i}}^2(\theta,\phi)\sin \phi
d\theta d\phi}\label{eq:G3Dz}
\end{equation}
\end{widetext}

Here $\tau_{\mathrm{N}z,\,\mathrm{i}}(\theta,\phi)$ is obtained as
in equation \ref{eq:tau_3Dx} but by projecting onto the $\mathbf{k}$
direction.

It is trivial to show that, in the presence of a single band and
when not only the FS in N but also the one in S is spherical,
equations  \ref{eq:G3Dx} and \ref{eq:G3Dz} reduce to those of the 3D
BTK model already discussed at the beginning of the present
subsection.

\begin{figure}[ht]
  \includegraphics[width=0.7\columnwidth]{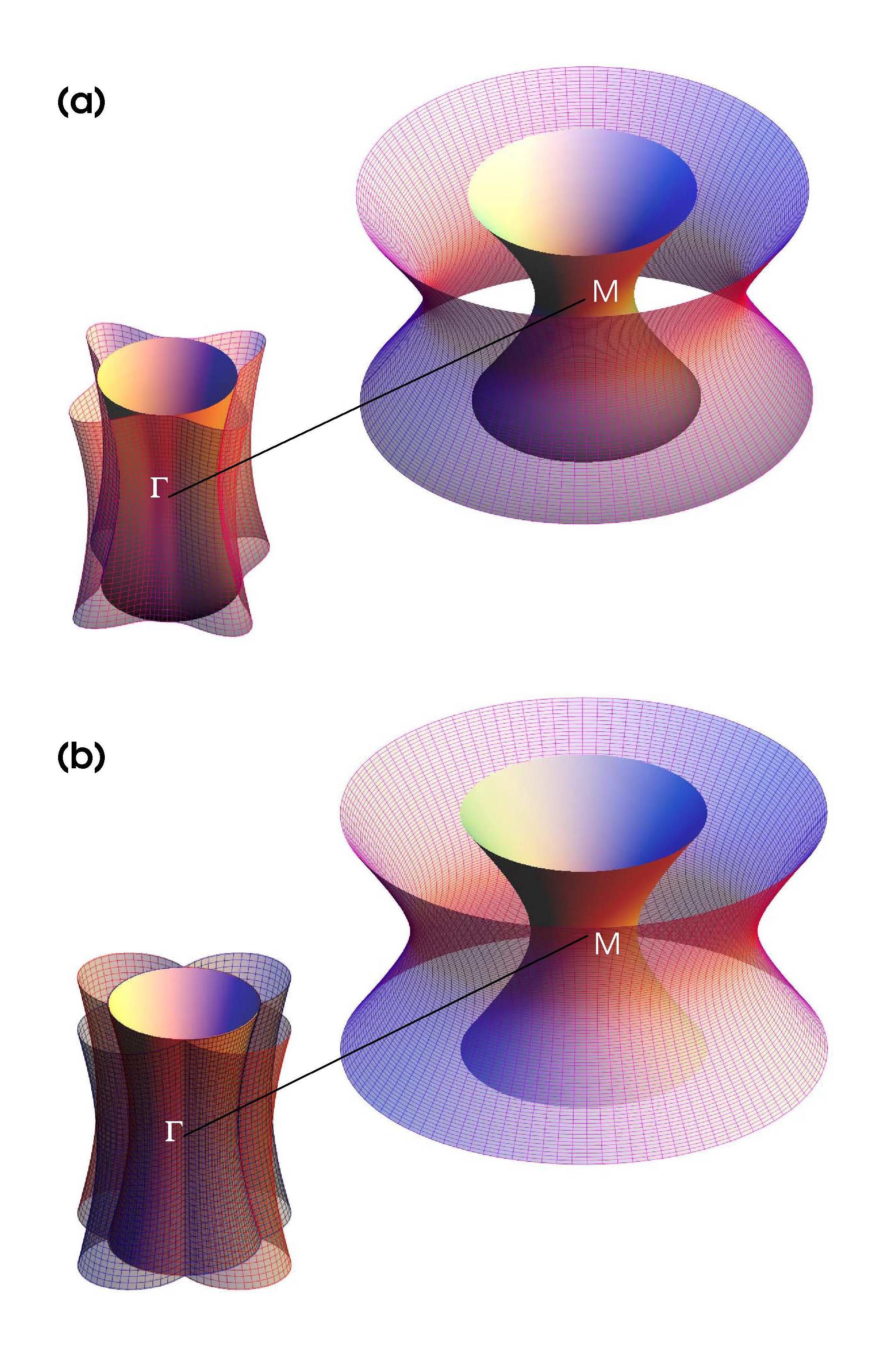}\\
  \caption{(color online): (a) Reciprocal space representation of the
shape of the FS sheets (matt surfaces) and of the pair potentials
(semi-transparent gridded surfaces) used in the calculations whose
results are shown in Fig. 4 (see text). In this case the OP on the
electron FS sheet at M has isotropic \emph{s}-wave symmetry while
the one on the hole FS sheet at $\Gamma$ has a fully anisotropic
\emph{s}-wave\emph{ }symmetry; (b) The same as in panel (a) but for
a \emph{$d_{\mathrm{x}^{2}-\mathrm{y}^{2}}$} symmetry of the OP on
the hole FS sheet at $\Gamma$.}\label{fig:3}
\end{figure}

We can now discuss the results of these calculations in a couple of
cases of particular interest for Fe-based superconductors. In the
typical 1111 or 122 compounds, the FS sheets can be approximated by
more-or-less warped cylinders \cite{mazin09}. In some special cases,
3D pockets have been calculated, that can instead be approximately
described as ellipsoids \cite{ivanovskii11}.

Surfaces that approximate reasonably well these FS sheets and that
can be easily written in parametric form are the oblate or prolate
spheroids and the one-sheeted hyperboloids of revolution. Here we
will use two hyperboloids with different (negative) Gaussian
curvatures to represent the hole and electron FS sheet of a
hypothetic 1111 or 122 Fe-based compound. These theoretical FS
sheets are shown as matt surfaces in figure \ref{fig:3}(a) and (b).
As far as the symmetry of OPs is concerned, we will consider here a
isotropic \emph{s}-wave pair potential on the electron FS sheet at
M, and a \emph{fully anisotropic s}-wave symmetry with zeros (figure
\ref{fig:3} (a)) or a $d_{\mathrm{x}^{2}-\mathrm{y}^{2}}$-wave
symmetry (figure \ref{fig:3} (b)) for the hole-like FS sheet at
$\Gamma$. In figure \ref{fig:3} the OPs are represented as
semi-transparent gridded surfaces with red grids where the pair
potential is positive and blue ones where it is negative.

\begin{figure}[ht]
  \includegraphics[width=\columnwidth]{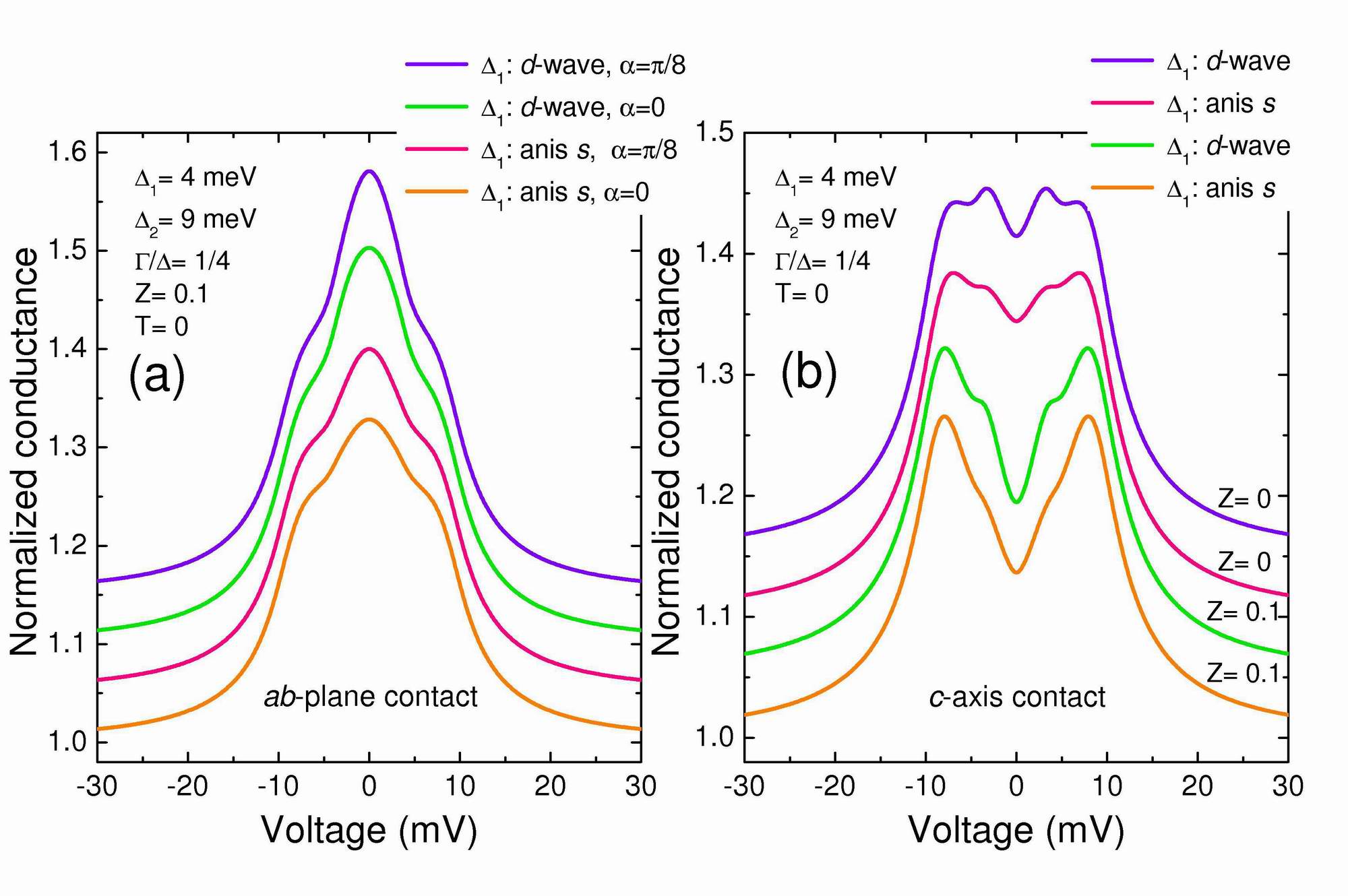}\\
  \caption{(color online): (a) Normalized AR conductance calculated by
using the full 3D BTK model described in the text for the FS shapes
and the two OP symmetries shown in Fig. 3, for two different values
of $\alpha$ angle and for current injection along the
\emph{ab}-plane; (b) The same as in panel (a) but for current
injection along \emph{c}-axis and for two different values of $Z$.
The other parameters of the calculations are shown in the labels and
discussed in detail inside the text.}\label{fig:4}
\end{figure}

By using these models for the geometry of FS sheets and symmetry of
OPs and by working out equations \ref{eq:G3Dx} and \ref{eq:G3Dz} we
obtain the theoretical normalized conductance for point contacts
along the \emph{x}-axis direction (\emph{ab}-plane contacts) and
along the \emph{z}-axis one (\emph{c}-axis contacts). This
conductance is shown as a function of bias voltage at $T=0$ and for
different values of the parameters in panels (a) and (b) of figure
\ref{fig:4}, respectively. In all the simulations we used:
$\Delta_{1}=4$ meV (maximum absolute value of the anisotropic pair
potential), $\Delta_{2}=9$ meV (value of the isotropic pair
potential), $\Gamma_{1}/\Delta_{1}=\Gamma_{2}/\Delta_{2}=1/4$, and
$Z_{1}=Z_{2}=0.1$. In practice the small gap is on the hole FS and
is anisotropic, with line nodes or zeros along four specific
directions, while the large one is on the electron FS and is
isotropic. In addition, $Z$ values similar to the experimental ones
and reasonably small (but not null) broadening parameters are used.
The selected symmetries and parameters do not correspond to any
actual Fe-based superconductor. Nevertheless, they are somehow
inspired by recent theoretical papers on hole-doped
$\mathrm{BaFe_{2}As}_{2}$ \cite{graser10} and on
$\mathrm{BaFe_{2}(As_{1-x}P_{x})_{2}}$ \cite{suzuki11} where 3D
nodal structures are predicted to appear in the warped hole FS.
Symmetries involving nodes on the electron FS (nodal \emph{s}-wave)
or on the hole FS (\emph{d}-wave) have been discussed also for 1111
low- $T_c$ compounds \cite{kuroki09}. Figure \ref{fig:4}(a) shows
the normalized conductance curves of \emph{ab}-plane contacts for
the two selected symmetries of the small gap and two different
values of the $\alpha$ angle. It is clearly seen that, independently
of the anisotropic symmetry of the small OP -- provided it has zeros
in some directions -- and of the injection angle with respect to
these directions, the \emph{ab}-plane curves present a more or less
pronounced zero-bias peak due to the small gap $\Delta_{1}$ and
broader shoulders originated by the isotropic gap $\Delta_{2}$. This
result is somehow expected if one considers the curves shown in Fig.
2 for small $Z$ values and $\Gamma\neq0$. Quite different is the
situation of \emph{c}-axis contacts. As shown in Fig. 4 (b) the
normalized conductance in this case presents no zero-bias peaks,
independently of the symmetry of the small OP. On the whole, the
curves appear very similar to what is expected in \emph{s}-wave
symmetry but for enhanced $Z$ values (of the order of 0.2-0.4). Also
the structures due to the small OP have a \emph{s}-like appearance
in both the symmetries and even for $Z=0$, always resulting in more
or less pronounced shoulders or peaks at about 4 meV in the total
conductance. This result is only partially expected. In fact now the
HLQs and the ELQs feel the same phase independently of $\theta$ and,
thus, interference effects due to phase differences cannot occur.
Nevertheless, the full integration in $0<\theta<2\pi$ gives rise to
an average of conductance curves representative of the different gap
values that are seen at different directions in the \emph{ab} plane.
As it happens in the \emph{ab}-plane contacts of panel (a), the
contribution of arbitrary small OPs close to the nodes or zeros
should yield zero-bias maxima in the total conductance. But this is
not the case. The reason is an additional and pure geometric effect.

When the current is injected along a direction mainly parallel to
the FS (as in this case, when the current is injected along the $c$
axis) the normal transparency
$\tau_{\mathrm{N}z,\,\mathrm{i}}(\theta,\phi)$ is markedly reduced
in amplitude at any angle and its overall shape is modified
accordingly. It can be easily shown that both these effects can be
roughly simulated by inserting a properly enhanced $Z$ value into
the
$\tau_{\mathrm{N}z}(\theta,\phi)=\cos^{2}\phi/\left(\cos^{2}\phi+Z^{2}\right)$
valid for a spherical FS. For the FS geometries here discussed this
$Z$\emph{-enhancing }effect occurs also in \emph{ab}-plane contacts
but to a quite minor extent since, in this case, most of the FS is
almost perpendicular to the direction of current injection. In other
words we can roughly say that, the larger is the projection of the
FS along the tunneling direction, the smaller is the $Z$-enhancing
effect. Of course, as extreme limits we have the spherical FS (where
the $Z$ enhancing is null for every tunneling direction) and the
perfectly cylindrical FS (along whose axis the enhanced $Z$ tends to
infinity and, thus, $\tau_{\mathrm{N}z}(\theta,\phi)$ tends to
zero). In the latter case, however, the resulting normalized
conductance is zero as expected from equation \ref{eq:G3Dz} and
already pointed out in \cite{daghero10}. As a consequence of this
$Z$-enhancing effect related to the FS-geometry, of the strong $Z$
dependence of the zero-bias peaks shown in figure \ref{fig:2} and of
the absence of phase-difference interference effects, it turns out
that \emph{c}-axis contacts on superconductors with warped
cylindrical FSs cannot show zero-bias conductance peaks or maxima
even in the presence of \emph{d}-wave or \emph{fully anisotropic
s}-wave symmetry of the OPs. In this case only zero-bias peaks or
maxima in \emph{ab}-plane contacts can give information on the
anisotropic symmetry of one or more of the OPs but, as shown in
figure \ref{fig:4}(a), without allowing the complete determination
of this symmetry. In section \ref{sect:PCAR} we will provide a
couple of examples in which zero-bias maxima have been observed
experimentally in single crystals and polycrystals of Fe-based
compounds of different families.

These results are also interesting from another point of view. By
reversing the previous reasoning one can say that if the OP on a FS
sheet has a \emph{d}-wave or \emph{fully anisotropic s}-wave
symmetry and AR experiments in \emph{c}-axis direction do show
zero-bias peaks or maxima, it means that the $Z$-enhancing effect is
negligible, i.e. this FS sheet has a large projected area on the
$xy$ plane. In materials layered along the \emph{ab} plane, this
occurs when the FS sheet has a closed surface, i.e. when it is
three-dimensional. This situation can be easily simulated by
substituting one of the two FSs of figure \ref{fig:3} with an oblate
or prolate spheroid, but a detailed discussion of these results is
beyond the scope of the present paper. The only important message we
would like to leave here is that in some particular situations, e.g.
in presence of anisotropic pair potentials, \emph{directional}
point-contact AR spectroscopy can provide information not only on
the \emph{symmetry of the OPs} but also on the \emph{geometry of FS
sheets} where the superconducting gaps open. A less univocal and
drastic, but still interesting information on the FS geometry comes
in any case from the observation that in the framework of this full
3D BTK model the weight factor of every band in the total normalized
conductance comes directly from the FS topology and the
directionality of the contact and \emph{is not} a fit parameter as
in the standard two-band fitting approach. It means that once the
geometry of FS sheets is fixed (e.g. by first-principle
band-structure calculations) it is then possible to compare the
experimental PCAR results with the theoretical 3D BTK predictions
getting information on the true FS topology as compared to the
theoretical one. Just as an example, we report here the weight
factors of bands 1 and 2 coming from the results shown in figure
\ref{fig:4} and, thus, for the geometries shown in figure
\ref{fig:3}. They are: $w_{1}^{ab}=0.48$, $w_{2}^{ab}=0.52$,
$w_{1}^{c}=0.33$ and $w_{2}^{c}=0.67$. These numbers are not far
from the actual values obtained in PCAR experiments on Fe-based
superconductors, particularly as far as $w_{i}^{ab}$ is concerned
\cite{gonnelli09a,daghero09b,tortello10}.

In principle, one should always use this full 3D model in fitting
the experimental data. However, i) the 3D model is rather
complicated to handle as a fitting tool; ii) it requires the
knowledge of the FS geometry, which is not always available; iii)
most of the Fe-based superconductors have quasi-2D Fermi surfaces so
that the 2D model can be safely used as a first approximation.
Indeed, the difference between the 3D and the 2D models is dramatic
only along the c axis, and in the case of \emph{d}-wave or
anisotropic \emph{s}-wave symmetry as shown in figure \ref{fig:4}.
In the case of \emph{s}-wave symmetry it is possible to show that
the results obtained with the 2D model are totally compatible with
those of the full 3D one. For example, we calculated the normalized
conductances for injection along the \emph{ab} plane and along the
\emph{c} axis by using the full 3D model, using a 3D FS made of two
moderately-warped one-sheeted hyperboloids of revolution. The two
\emph{s}-wave gaps and the other parameters used in the calculation
are the same as in figure \ref{fig:4}, except the barrier heights
that here are $Z_1=Z_2=0.2$. When the theoretical 3D conductance
curves are fitted by the 2D BTK model the resulting parameters are
perfectly coincident with the starting ones (including the weights
of band 1 and 2) with only the expected exception of the parameter
$Z$ that results about 20\% larger along \emph{ab} plane and 2.8-3.6
times larger than the actual one in the fit of the \emph{c}-axis
conductance, due to the mentioned $Z$-enhancing effect. As a
consequence, in \emph{s}-wave symmetry, the use of the fully 3D
model is mandatory only if a precise determination of the $Z$
parameters is required.

In this rather long subsection we have discussed in detail the
effects of FS geometry and directionality of the contact on the
results of point-contact AR spectroscopy in multiband multi-gap
superconductors like the Fe-based ones. In the next subsection we
will focus instead on the effect in PCAR curves of the presence of a
strong electron-boson interaction and on the role played by the
shape of the electron-boson spectral function, by appropriately
solving a three-band $s\pm$ Eliashberg model for the determination
of the $\Delta_{\mathrm{i}}(E)$ functions to be inserted in the
above reported expressions for the normalized conductance.

\subsection{Electron-boson interaction and shape of the spectral function}\label{subsect:EBI}

It is well known that, in the BCS theory, the OP does not depend on
energy. This is not the case when the Eliashberg theory of
superconductivity is adopted. By starting from the electron-boson
spectral function $\alpha^2F(\Omega)$ it is indeed possible, within
the Eliashberg approach, to account for the full energy dependence
of the OP which is now described by a complex function of energy,
$\Delta(E)=\Re \Delta(E) +i\Im\Delta(E)$. In particular, as the
strength of the coupling increases, the imaginary part also
increases and small deviations from the BCS density of states can be
observed at the typical boson energies. In the
moderate/strong-coupling regime these deviations can be
experimentally detected by tunnelling and PCAR spectroscopy.

\begin{figure*}[t]
\begin{center}
\includegraphics[keepaspectratio, width=\textwidth]{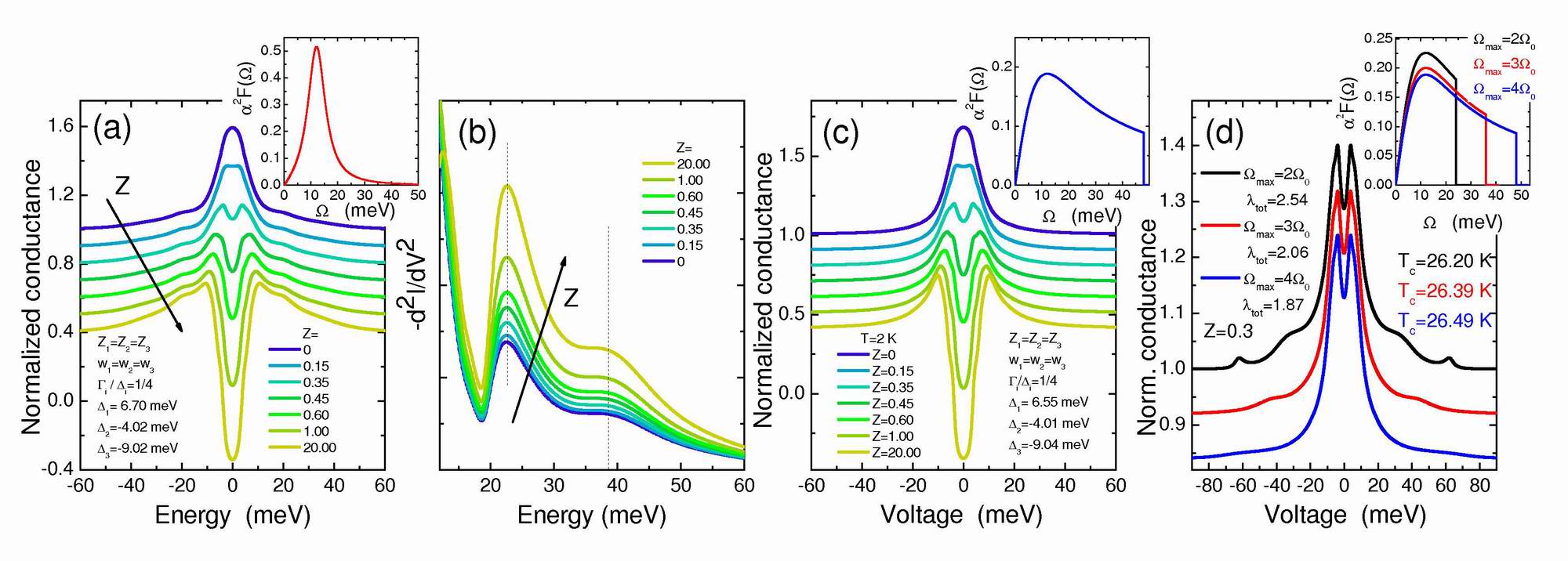}
\end{center}\caption{(color online) (a): Andreev-reflection spectra
obtained with different Z values by introducing in the BTK model the
energy-dependent gaps calculated within a three-band $s\pm$
Eliashberg model and by using a Lorentzian $\alpha^2F(\Omega)$
(inset) with $\Omega_0=12$ meV and half width $\Upsilon=4$ meV. The
coupling constant $\lambda_{tot}=2.11$ has been fixed so that $T_c =
24.92$ K. The gaps given by the Eliashberg calculations are
$\Delta_1=6.70$ meV, $\Delta_2=-4.02$ meV and $\Delta_3=-9.02$ meV
and correspond to the case of Ba(Fe$_{0.9}$Co$_{0.1}$)$_2$As$_2$
\cite{tortello10}. The other parameters of the BTK model are
$Z_i=0.3$, $\Gamma_i/\Delta_i=1/4$ and w$_i=1/3$. The sign-changed
derivative of the conductance curves are shown in (b). (c): same as
in (a) but using a AFSF spectral function (inset) with $\Omega_0=12$
meV, $\lambda_{tot}=1.87$ and $\Omega_{max}=4\Omega_0$. Calculated
$T_c$ is 26.49 K. The gaps given by the Eliashberg calculations are
$\Delta_1=6.55$ meV, $\Delta_2=-4.01$ meV and $\Delta_3=-9.04$ meV.
The other parameters of the BTK model are Z$_i=0.3$,
$\Gamma/\Delta_i=1/4$ and w$_i=1/3$. (d):  Conductance curves
calculated as in (c) but for $Z=0.3$ and different values of
$\Omega_{max}$ in the AFSF $\alpha^2F(\Omega)$. From bottom to top,
$\Omega_{max}=4\Omega_0$, $\Omega_{max}=3\Omega_0$ and
$\Omega_{max}=2\Omega_0$. The three spectral functions are shown in
the inset.}\label{Fig5}
\end{figure*}

From the theoretical point of view, a precise knowledge of the
shape of the spectral function is not really necessary, at least
to a first approximation, to calculate $T_c$, the gaps etc. within
the Eliashberg theory, since these quantities mostly depend on the
representative frequency of $\alpha^2F(\Omega)$
\cite{nicol05,ummarino04,valerogiannis95}. However, we will show
in the following that the shape of the spectral function and the
maximum boson energy affect the possibility to observe
electron-boson interaction features in the Andreev-reflection
spectra. In this regard, we will compare different shapes of the
spectral function, namely a Lorentzian function \cite{ummarino09}
(see inset to figure \ref{Fig5}a) and the typical
antiferromagnetic spin fluctuations (AFSF) spectral function
\cite{bose03} (inset to figure \ref{Fig5}(c)).

The effect of the shape of the spectral function on the
electron-boson features will be discussed within a three-band,
s$\pm$ Eliashberg model which revealed useful in describing several
Fe-based superconductors \cite{ummarino09,ummarino10b,tortello10}.
Within this model, the electronic structure of Fe-based
superconductors is described by two hole bands and one equivalent
electron band in the case of hole doping, and by two electronic
bands and one equivalent hole band in the case of electron doping.
The other main assumptions of the model are that: i) the total
electron-phonon coupling constant is small \cite{boeri08,boeri10};
ii) phonons mainly provide intraband coupling; iii) spin
fluctuations mainly provide interband coupling. We will choose, as a
representative example, the compound
Ba(Fe$_{0.9}$Co$_{0.1}$)$_2$As$_2$ \cite{tortello10}. The parameters
($\Delta$, $Z$, $w$ etc.) will be indexed according to the bands,
i.e. 1 for the hole FS, 2 and 3 for the electron FS sheets. Further
details on the model and on the parameters of the calculations can
be found in \cite{tortello10}.

Figure \ref{Fig5}(a) shows several Andreev-reflection conductance
curves calculated by using a three-band 2D BTK model (equation
\ref{eq:G2D}) in which, however, the OPs are not constant but
energy-dependent. The functions $\Delta_i(E)$ were previously
calculated within the aforementioned three-band Eliashberg model by
using coupling constants able to give $T_c$ and gaps similar to the
experimental ones. The $\alpha^2 F(\Omega)$ was chosen to have a
Lorentzian shape, with representative boson frequency $\Omega_0=12$
meV, similar to the spin-resonance energy obtained from neutron
scattering experiments in $\mathrm{Ba(Fe_{1-x}Co_x)_2As_2}$
\cite{inosov10}. The curves in figure \ref{Fig5}(a) refer to
different values of the Z parameters ($Z_1=Z_2=Z_3=Z$) from 0 (ideal
AR contact with no barrier) to 20 (tunnel regime). Each conductance
curve is the sum of three contributions (one for each band of the
model) whose relative weight have been here chosen to be equal,
$w_1=w_2=w_3=1/3$. The other parameters of the BTK model have been
chosen so as to compare to real experimental situations, i.e. $T=2$
K, $Z_1=Z_2=Z_3=0.3$ and $\Gamma_i/\Delta_i=1/4$. The calculated
conductance curves in figure \ref{Fig5}(a) feature structures at
energies higher than the gap values, at about
$\Delta_{max}+\Omega_0=21$ meV, which become more pronounced as $Z$
increases. The features can be seen much more clearly by looking at
the sign-changed derivative of the conductance, $-dG/dV=-d^2I/dV^2$
shown in figure \ref{Fig5}(b). A peak is present above 20 meV and an
additional feature can be also seen around 40 meV (vertical dashed
lines). When $Z$ decreases, the structures slightly reduce in
amplitude but their position in energy hardly changes, showing that
important information concerning electron-boson interaction can be
obtained in the Andreev-reflection regime as well.

Figure \ref{Fig5}(c) shows the same cases as in (a) but now
calculated by adopting a normal AFSF spectral function, with maximum
at $\Omega_0 = 12$ meV and maximum boson energy
($\Omega_{max}=4\Omega_0=48$ meV). Here the total coupling constant
$\lambda$ is fixed so as to give (approximately) the same $T_c$ and
the same gaps as in panel (a). The additional structures are no
longer visible in the conductance curves at any value of the $Z$
parameter. Only tiny deviations from the BCS case can be seen in the
sign-changed derivative (not shown) but these features would be
hardly observable in a real experiment. Figure \ref{Fig5}(d) shows
what happens if the value of $\Omega_{max}$ is reduced, though
scaling $\lambda_{tot}$  so as to give the same $T_c$ and gaps (see
the inset). If $\Omega_{max}=3\Omega_0$ small additional structures
start to become visible and they grow consistently in amplitude when
$\Omega_{max}=2\Omega_0$, but now they are at
$\Delta_{max}+\Omega_{max}$ and not at $\Delta_{max} + \Omega_0$.
Summarizing, in this case the electron-boson interaction structures
are detectable only if the spectral function is abruptly cutoff
resulting in a shape no more physically plausible (see inset to
figure \ref{Fig5}(d)) and are located at energies that have nothing
to do with the representative boson energy $\Omega_0$. The
conclusion is that reasonable parameters for the normal AFSF
spectral function do not allow reproducing features that are
actually observed in the experiment (see section
\ref{sect:EBIinFebased}). Hence, we suggest that this particular
shape of the $\alpha^2F(\Omega)$ is not suitable to catch finer
details of the PCAR spectra measured in these compounds. This fact,
however, may not be surprising since the AFSF function used here was
calculated for a uniform electron gas in the normal state
\cite{bose03} which looks too rough an approximation in this case.
Very recently, the AFSF electron-boson spectral function in the
superconducting state has been calculated \cite{nagai11} and is
indeed rather similar to the Lorentzian $\alpha^2F(\Omega)$ used
above, which is thus a reasonable and easy-to-handle approximation.

\begin{figure} [h]
\begin{center}
\includegraphics[keepaspectratio, width=1 \columnwidth]{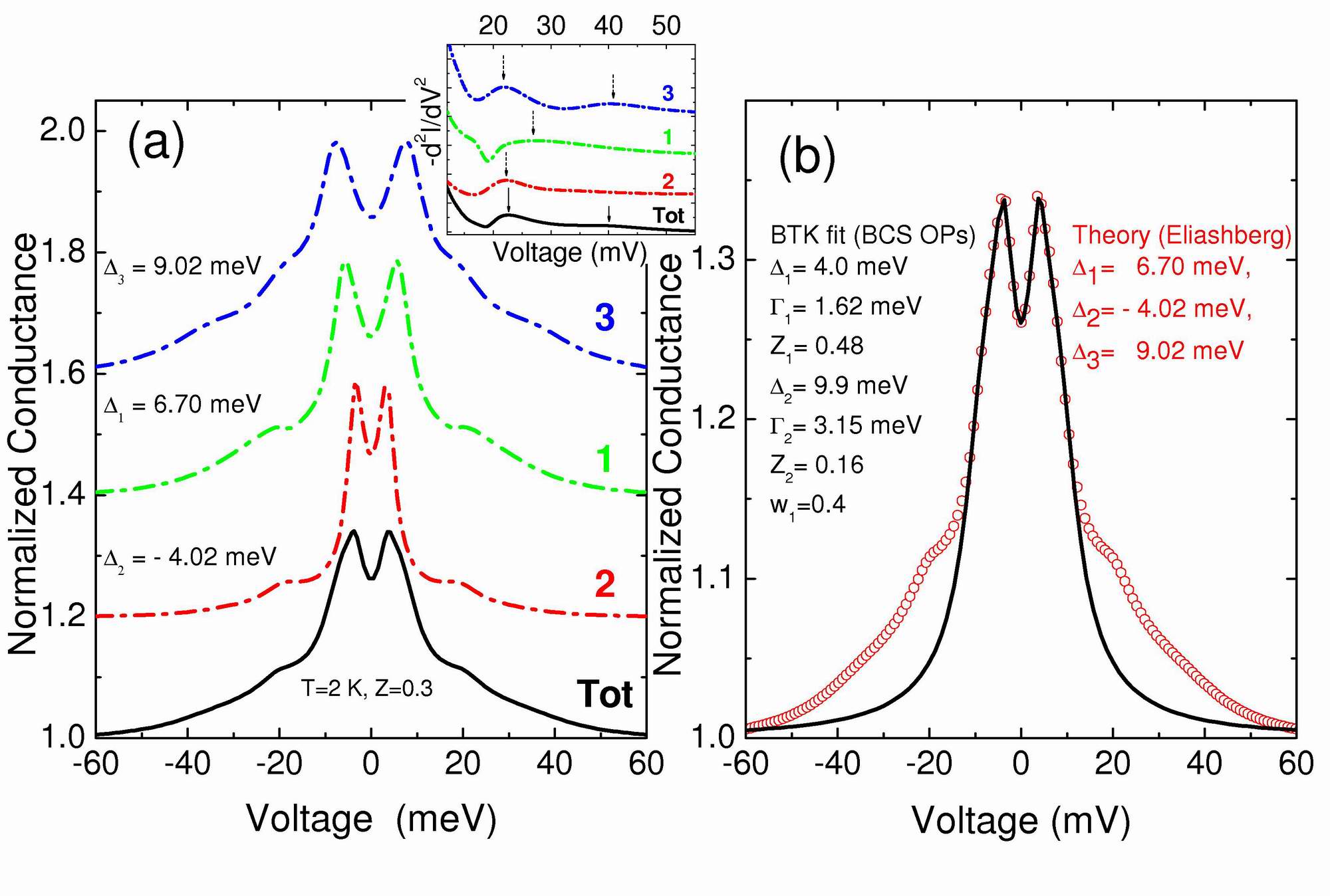}
\end{center}
\caption{(color online): Solid line: total conductance curve
calculated with the same parameters of figure \ref{Fig5}(a) but
with Z=0.3. Dash-dot lines: single band contributions to the total
conductance. Inset: sign-changed derivatives of the curves shown
in the main panel. (b) circles:  total conductance curve from
panel (a) compared to the 2-gap 2D BTK fit (solid line) with
constant BCS values for the gaps. The parameters of the fit are
reported in the left label.}\label{Fig6}
\end{figure}

Figure \ref{Fig6}(a) shows an Andreev-reflection spectrum (solid
line) calculated at $T\!=\!2$ K, $Z\!=\!0.3$ and
$\Gamma_i/\Delta_i=1/4$ by using a Lorentzian spectral function,
together with the partial contributions from the three bands
present in the model (dash-dot lines). Each of the three partial
conductance curves shows strong-coupling structures at energies
higher than the gap. However, while in a single-band case one
could expect these features to appear at $\Delta+\Omega_0$ (and
potentially at higher harmonics), in a three-band case with a
dominant interband pairing the situation is not so
straightforward. For instance, the contribution of band 2 features
a peak at $\Delta_3+\Omega_0\approx 22$ meV and not at
$\Delta_2+\Omega_0$. This fact, which is most probably due to the
strong interband character of the coupling, has to be investigated
further and suggests that quantitative calculations are
recommended in order to reliably interpret electron-boson features
in Fe-based superconductors.

Figure \ref{Fig6}(b) shows what happens when the total conductance
of panel (a), here represented by circles, is fitted to the two-band
2D BTK model which adopts constant BCS energy gaps (solid line). The
agreement between the two curves is remarkable in the gap region,
despite the fact that the BTK model contains only \emph{two} gaps (a
three-band model would contain too many parameters for the fit to be
reliable). The fit allows determining very well the small gap (4.0
meV against 4.02 meV, see the parameters in the labels). The large
gap obtained by the fit is similar to $\Delta_3$ given by the
Eliashberg theory, with a slight overestimation (9.9 meV against
9.02 meV); the effect of the intermediate gap is a hardly visible
excess of signal with respect to the fitting curve between the
extremal gaps, which is indeed sometimes observed experimentally. At
higher energies a considerable deviation from the fit is observed:
The calculated conductance shows additional features together with a
higher conductance up to about 60 meV. Such an excess of conductance
at intermediate energy has been observed rather often by PCAR in
Fe-based superconductors (see figure 2d and 2e from \cite{chen08b}
or figure \ref{Fig13}(a) below). Furthermore, a higher-energy excess
conductance that some calculated curves show for particular bands
can cause an increase of the amplitude of the tails even up to
80-100 meV, as shown in figure \ref{Fig13}(a) and in the inset to
figure \ref{fig:yates}(a). This effect can be a further evidence of
the strong-coupling character of these compounds. However, since the
height of the tails at high energy in the experimental curves
depends on the normalization procedure, one must be confident that
the actual normal state is adopted, i.e. the curve at low
temperature and $H>H_{c2}$. Since the upper critical field in these
compounds is usually not experimentally accessible, one could try
using the normal state at $T>T_{c}$, provided that no heating
effects are present in the point-contact which could cause a
downward shift of the conductance curves with increasing
temperature. Therefore, a systematic study of this effect is
desirable, both from the experimental and theoretical point of view,
as it could greatly help acquiring more information concerning the
strength of the coupling and the energy of the mediating boson in
Fe-based superconductors.

The fact that in the above discussion we have compared the same data
to either a two-band BTK model with constant (energy-independent)
OPs or a three-band BTK model with energy-dependent OPs might look a
little confusing. Let us just explain this point.

One one hand, if one starts from an experimental PCAR spectrum, the
\emph{minimal} BTK model able to fit the data -- at least in the
small-to-intermediate energy range -- is the two-band BTK one. In
some spectra, a discrepancy between the data and the fitting curve
is observed at energies between the small and the large gap, which
might be due to the presence of a third gap, which is however
experimentally difficult to discern. Moreover, this discrepancy is
too small to justify the use of a three-band BTK model that would
contain 11 free parameters and, thus, would not be reliable. In this
sense, the large uncertainty on the gap $\Delta_2$ (especially in
the case of Sm-1111) might partly arise from the oversimplification
of the fitting model. Even if this fact has not prevented obtaining
interesting information by PCAR measurements, this consideration has
to be kept well in mind when studying these complex multiband
systems.

On the other hand, the \emph{minimal} Eliashberg $s\pm$ model that
allows obtaining the experimental gaps and $T_{c}$ is a three-band,
three-gap one as explained in \cite{ummarino09}. If one wants to
take into account the effects of the energy dependence of the OPs
and of the strong electron-boson coupling on the shape of the PCAR
spectrum, one is thus forced to generate the Andreev-reflection
curve by using a three-band BTK model with the three OPs coming from
Eliashberg calculations. Note that, in the latter case, the OPs are
not fitting parameters and only $\Gamma_i$, $Z_i$ and the weights
can be adjusted to ``fit'' the experimental spectrum.

\section{Symmetry of the order parameters in Fe-based
superconductors}\label{sect:PCAR}

\subsection{``1111'' compounds}
In most of the 1111 compounds, the model for spin-fluctuation
mediated superconductivity predicts a $s\pm$ gap symmetry with a
sign change between hole-like and electron-like Fermi surface
sheets. This is indeed compatible with what has been found by
point-contact measurements, despite initial contradicting claims.
An order parameter with nodes on the electron FS has been
predicted in 1111 compounds with low $T_c$ \cite{kuroki09} which
have not been studied yet by PCAR.

The first PCAR measurements in 1111 materials like LaFeAs(O,F) and
SmFeAs(O,F) showed zero-bias peaks that were initially interpreted
as being due to a nodal gap symmetry, namely the $d$-wave one.
However, the appearance of these zero-bias features and their height
were later discovered to depend on the pressure applied to the
sample by the tip (in the needle-anvil configuration)
\cite{yates08a}; in some samples, they were frequently found in some
regions of the sample surface and never in others; moreover, their
temperature dependence was odd and completely inconsistent with that
expected for the zero-bias peak associated with Andreev bound states
in nodal superconductors \cite{samuely09a,yates09}. At the same time
and most importantly, increasing pieces of evidence were collected
that, in suitable conditions (better sample quality, or zero
pressure applied to the sample as in the ``soft'' technique
described in \cite{daghero10}) the zero-bias peak did not show up
\cite{chen08b,gonnelli09a,daghero09b}. In the end, the majority of
the scientific community got convinced that the zero-bias peaks
early observed in 1111 compounds were not a signature of $d$-wave
symmetry of the order parameter. We will thus not discuss these
results any longer; a short review on this subject can be found in
\cite{daghero10}.

The absence of zero-energy bound states (particularly in $ab$-plane
contacts) certainly excludes those symmetries where nodal lines lie
on the Fermi surface. However, the predicted symmetry of the OP in
pnictides does involve a change of sign in the OP, but this should
occur \emph{between} hole-like and electron-like Fermi surface
sheets \cite{mazin08}. According to some models, this sign change
should give rise to observable effects in the PCAR spectra through
interband interference effects. The article by Greene et al. in this
same issue is specifically devoted to exploring these models and
comparing their predictions to experiments. Here we will not deal
with this aspect and show instead how the most reliable PCAR spectra
in 1111 can be successfully fitted by simpler models that disregard
interference effects. Indeed, the predicted features associated with
interband interference (i.e. surface bound states, both at zero and
nonzero energies \cite{golubov09}) may be often made indiscernible
by various sources of broadening or overwhelmed by stronger effects.

It is however worth mentioning that, in contacts with vanishing
barrier at the interface (let's say, $Z=0$ in the BTK language), a
strong reduction in the Andreev signal is predicted \cite{golubov09}
as a result of destructive interference.  In Fe-based
superconductors a smaller amplitude of the Andreev signal than in
more conventional systems (even multiband, like MgB$_2$) is indeed
often (but not always) observed experimentally. However, a similar
reduction in the amplitude of the superconducting signal is also
expected if the normal bank of the contact is a diffusive metal
\cite{tanaka03,shigeta08}, if a large inelastic scattering occurs at
the interface \cite{chalsani07}, if a fraction of the probe current
is injected in normal regions (as it can happen if the dopant
distribution is inhomogeneous), or even if some Fermi surface sheets
are gapless. For the sake of completeness, it must be said that a
small Andreev signal is also predicted if the gap is anisotropic
(see figure \ref{fig:2}). At this stage of the research it is
impossible to say which of these effects is really taking place.
This is particularly true for 1111, since the growth of large single
crystals is still very difficult and most of the experiments have
been carried out on polycrystalline samples which necessarily do not
allow ruling out extrinsic effects due to grain boundaries, surface
degradation, inhomogeneous doping and so on.

\subsubsection{LaFeAs(O$_{1-x}$F$_x$).}

LaFeAs(O$_{1-x}$F$_x$) is chronologically the first Fe-based
superconductor studied by point-contact spectroscopy, and also the
first on which we performed ``soft'' PCAR measurement (where the
contact is made by placing a small drop of Ag paste on the sample
surface instead of pressing a sharp metallic tip against it) to
avoid pressure effects that could possibly give rise to spurious
zero-bias peaks, as shown in some early measurements reported in
literature.\\

\begin{figure*}
\includegraphics[width=\textwidth]{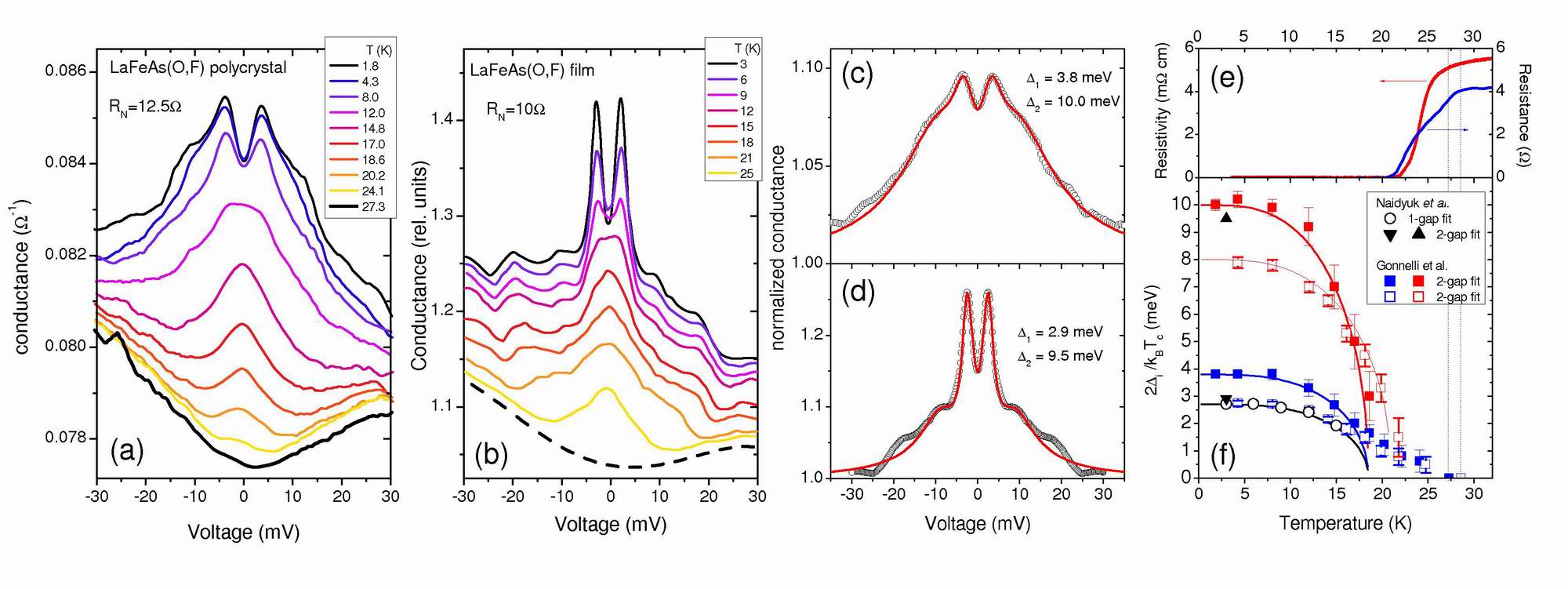}\\
\caption{(color online): Point-contact spectra in
LaFeAsO$_{1-x}$F$_x$ polycrystals \cite{gonnelli09a} and films
\cite{naidyuk10}. (a, b): raw conductance curves of two contacts
with similar critical temperature $T_c^A$ obtained in polycrystal
and film, respectively. In (b), the normal-state conductance was
not measured and is simulated by the bottom dashed line. (c, d):
the low-temperature conductance curves from (a) and (b) after
normalization (i.e. division by the normal-state curve at
$T_c^A$), and the relevant two-gap 2D fit (lines). (e): comparison
of the resistivity of the polycrystal (left axis) to the
resistance of the film (right axis). (f): Temperature dependence
of the gaps as obtained from the 2-gap 2D fit of the curves in (a)
($\blacksquare$) and from the fit of the PCAR spectra of a
different contact, with $T_c^A=28.6$ K ($\square$), on the same
sample. The results of the 1-gap BTK fit of the curves in (b)
performed by Naidyuk \emph{et al.} is also shown ($\circ$) as well
as the 2-gap fit of their low-$T$ curve shown in (d)
($\blacktriangledown,\blacktriangle$). The lines are BCS-like
curves that fit the low-temperature trend of the gaps and, once
extrapolated, would give an ``apparent'' critical temperature
$T^*< T_c^A$.}\label{fig:La1111}
\end{figure*}

We studied polycrystalline samples with nominal F content $x=0.1$ and thus at optimal doping.
However, the local F concentration in the crystallites was found to
show deviations $\delta x = 0.02$ around this value. Indeed, point
contacts on different regions gave critical temperatures $T_c^A$
(defined as the temperature where the Andreev-reflection features
disappear and the normal-state spectrum is recovered) ranging from
27.3 K to 31.0 K. These temperatures lie in the upper part of the
resistive transition (see figure \ref{fig:La1111}(e)).

In most of the contacts we observed clear conductance maxima,
symmetric about zero bias and related to a small gap $\Delta_1$,
together with additional structures that we interpreted as being due
to a larger order parameter $\Delta_2$ (although we acknowledged the
possibility that they could be features of the normal state, which
is also changing with temperature \cite{gonnelli09a,arham11}). In
some contacts, these additional features were extremely clear and
sharp (see figure 2 in \cite{gonnelli09a}); in others (see figure
\ref{fig:La1111}(a)) they showed up as broad shoulders progressively
decreasing in height and width on increasing temperature. Other
peculiar features of the spectra are the right-left asymmetry, and
the residual upward curvature in the normal-state conductance that
looks like a ``pseudogap'' feature and is progressively reduced on
increasing the temperature. Curiously enough, the temperature at
which it disappears, leaving a flat but asymmetric conductance, is
close to the temperature of magnetic ordering in the parent
compound.

The origin of this behaviour is still not clear and will be possibly
clarified only when single crystals are available (see also
\ref{subsec Sm} for further discussion).

The complete absence of zero-bias peaks or maxima such as those
shown in figure \ref{fig:4} in the conductance, together with the
random orientation of the grains, made us conclude that none of the
order parameters shows line nodes or zeros on the Fermi surface. The
further conclusion in favor of a isotropic gap symmetry came instead
from the fit of the conductance curves, which was indeed always
possible by using two isotropic order parameters.

The fit requires of course a normalization of the conductance
curves, which is particularly crucial in these materials (as already
mentioned in section \ref{subsect:EBI}) because the gap or boson
structures can extend to very high energy. In some spectra, these
structures become clearly visible only if a proper normalization is
chosen. In principle, the normalization should be obtained by
dividing the superconducting conductance curve measured at a given
$T<T_c^A$ by the normal-state one at the same temperature. The
latter is experimentally inaccessible due to the large critical
fields of these materials \cite{hunte08}. Some authors then choose
to divide the raw conductance curve at a given temperature by a
straight line \cite{chen08b} that connects its high-energy
``tails''. This procedure generally allows a good estimation of the
small gap alone. As a matter of fact, the pseudogaplike curvature of
the normal-state conductance curves observed at $T>T_c^A$ is
presumably present also below $T_c^A$ (and likely to become more and
more pronounced when the temperature is lowered). This curvature is
completely disregarded by the aforementioned normalization
procedure. To partially correct for this, a possibility is to divide
all the raw conductance curves in the superconducting state by the
normal-state curve measured just above $T_c^A$, or (as we also did
in La-1111) to try to mimic the temperature dependence of the
(presumed) normal state \cite{gonnelli09a}. We showed that choosing
one or another of these two procedures changes very little the value
of $\Delta_1$ extracted from the fit (less than 2\%), while the
value of $\Delta_2$ can change by about 10\%, though preserving its
trend as a function of temperature, magnetic field and critical
temperature.

Figure \ref{fig:La1111}(c) shows the low-temperature experimental
curve of panel (a) after normalization (i.e. division by the normal
state at $T_c^A$), together with its best-fitting curve calculated
within the two-gap 2D BTK model discussed in section
\ref{subsect:BTK}. Let us just recall here that this model is really
suited to describe a 3D superconductor only if the gap is isotropic
and the Fermi surface is spherical, which is clearly not the case in
these materials \cite{singh08,mazin09}. More sophisticated models
accounting for the real shape of the Fermi surface (see section
\ref{subsect:FS}) should thus be used, but calculations may become
too heavy for a fitting procedure to converge. In practice, using an
approximated 2D model often simply results in an overestimation of
the barrier parameter $Z$, as already mentioned in section
\ref{subsect:FS}. The values of the OPs extracted from the fit of
the low-temperature conductance curve are $\Delta_1=3.8 \pm 0.1$ meV and
$\Delta_2=10.0 \pm 0.2$ meV. The fit of another set of curves measured
in a point contact with $T_c^A=28.6$ K, but featuring clearer
structures related to the second gap, gives instead $\Delta_1=2.75
\pm 0.1$ meV and $\Delta_2=7.9 \pm 0.2$ meV. How can these result be
reconciled with the $s\pm$ picture? As already mentioned above
(section \ref{subsect:EBI}) we have shown \cite{ummarino09} that the
$s\pm$ picture is compatible with the opening of gaps of different
amplitude on the two hole-like FS sheets and on the electron-like
sheet of hole-doped 1111 compounds. Eliashberg calculations assuming
a mainly-interband coupling due to spin fluctations give, for the
case of LaFeAs(O,F) analyzed here, three isotropic order parameters
of different signs: a small positive OP $\Delta_a= 2.82$ meV
(presumably on the outer hole FS cylinder around $\Gamma$), a large
positive OP $\Delta_b=8.01$ meV (on the inner hole FS cylinder), and
a large negative OP $\Delta_c=-7.71$ meV (on the electron FS sheets
around M) \cite{ummarino10b}. In the absence of interband
interference effects, Andreev reflection is only sensitive to the
amplitude (i.e. the absolute value) of the OPs (which is the actual
energy gap). This, together with the similarity between the absolute
values of the two large OPs, reconciles the findings of PCAR with
the theoretical expectations in the $s\pm$ model.

The gaps (or, better, the relevant $2\Delta/k_B T_c^A$ ratios)
extracted from the fit of the conductance curves in figure
\ref{fig:La1111}(a) are shown as solid squares in panel (f). Their
temperature dependence is anomalous; both $\Delta_1$ and
$\Delta_2$ follow a BCS-like trend up to $T^*\simeq 18$ K, but
above this temperature $\Delta_2$ is no longer detectable and
$\Delta_1$ shows a ``tail'' ending up at the real critical
temperature of the contact, $T_c^A=27.3$ K. The same result was
found in other contacts with different critical temperature
\cite{gonnelli09a}. The data shown by open squares in
\ref{fig:La1111}(f) refer to a contact with $T_c^A=28.6$ K on the
same sample; also in this case the large gap tends to zero at $T^*
\simeq 21$ K, and above this temperature the small gap shows a
``tail'' up to the real $T_c^A$.

Very similar results have been recently obtained by Naidyuk \emph{et
al.} \cite{naidyuk10} who performed PCAR spectroscopy with a Cu tip
in LaFeAsO$_{1-x}$F$_x$ films with $T_c = 28$ K (defined as the
temperature where the resistance drops to 90\% of its normal-state
value, see figure \ref{fig:La1111}(e)) and a transition width
$\delta T_c \approx 6$ K. The conductance curves $dI/dV$ obtained
from their data are shown in figure \ref{fig:La1111}(b), and show
two clear maxima related to a small gap $\Delta_1$. The authors
normalized the curves to a parabolic curve that fits the data for
$|V| \geq 10$ mV. This necessarily makes all the features at $|V|
\geq 10$ become small oscillations around 1 in the normalized curve.
As a matter of fact, the curves normalized in this way can be fitted
by a single-band, $s$-wave BTK model including a broadening term and
the fit thus gives only a small gap, shown in panel (f) as circles.
The gap follows a BCS-like curve that, once extrapolated, would give
again a critical temperature $T^*\simeq 18$ K. Structures in the raw
conductance curves, although impossible to fit with the same model,
persisted well above that temperature and at least up to 25 K, as
shown in panel b. As a matter of fact, the temperature dependence of
the raw conductance curves is not dissimilar from that observed by
us in polycrystals (panel a). Let us show what happens instead if
one normalizes the same curves by the normal state at $T_c^A$.
Although the corresponding curve is not reported in
\cite{naidyuk10}, based on the similarity with panel (a) one can
guess that it should look like the dashed curve in panel (b). If
this is used to normalize the low-temperature spectrum, one obtains
the curve depicted in panel (d), which shows a rather high Andreev
signal and shows clear shoulders in addition to the structures
related to $\Delta_1$. The two-band 2D model allows fitting this
curve very well, giving $\Delta_1 \simeq 2.9$ meV and $\Delta_2
\simeq 9.5$ meV (triangles in panel f). The agreement between the
data sets measured by us in polycrystals and by Naidyuk \emph{et
al.} in films is incredibly good if the raw conductance curves are
treated in the same way. This proves that: i) the different sets of
data are really compatible with each other; ii) even if in the raw
data by Naidyuk et al. the large-gap features are not as clear as in
ours, a suitable normalization makes them show up clearly enough to
be fitted by the 2D model; iii) the gap values are rather robust
against the sample form and synthesis technique, provided that the
critical temperature is similar, as in this case. The two data sets,
however, also pose some questions on whether the observed anomalous
tendency of the gaps to close around $T^*$ is an artifact or is
intrinsic to this material. The simplest possibility, of course, is
that the behavior of the gap is the result of the shortening of the
mean free path on increasing temperature, so that the junction
ceases to be ballistic at a voltage that decreases with temperature
\cite{daghero10}.

\subsubsection{SmFeAs(O$_{1-x}$F$_x$).} \label{subsec Sm}
\begin{figure*}
\includegraphics[width=\textwidth]{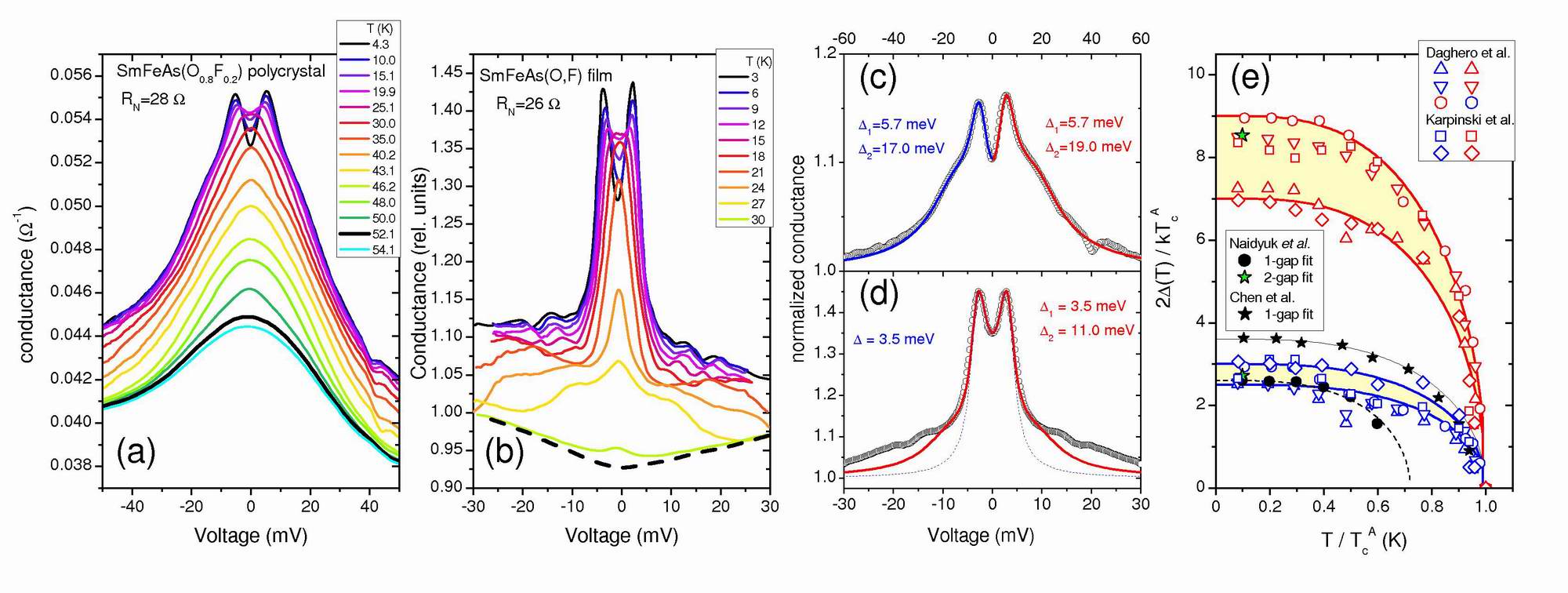}\\
\caption{(color online): PCAR spectra obtained in
SmFeAsO$_{1-x}$F$_{x}$ polycrystals \cite{daghero09b,gonnelli09b}(a)
and films \cite{naidyuk10}(b). The bottom curve in (b) is an
estimation of the normal state conductance. (c,d): the
low-temperature conductance curves from (a) and (b) normalized to
the normal-state conductance at $T_c^A$ and the relevant two-gap, 2D
BTK fit. In (c) the left and right side have been fitted separately;
in (d) the curve has been symmetrized. (e): Normalized gap values
($2\Delta/k_B T_c^A$) as a function of $T/T_c^A$ as obtained in
different samples: optimally doped polycrystals
($\vartriangle$,$\triangledown$, from \cite{daghero09b}); underdoped
polcrystals ($\circ$ from \cite{daghero09b}, $\bigstar$ from
\cite{chen08b}); single crystals ($\square$,$\diamondsuit$ from
\cite{karpinski09}); films ($\bullet$ from \cite{naidyuk10}). The
green stars indicate the results of the fit shown in panel (d).
Lines are BCS-like curves. The dashed line fits the trend of the gap
taken from \cite{naidyuk10}, which closes at the ``apparent''
critical temperature $T^*< T_c^A$. Colored regions approximately
indicate the region where the values of $\Delta_1$ and $\Delta_2$
are spread.}\label{fig:Sm1111}
\end{figure*}

Also in Sm-1111 the PCAR results reported in literature apparently
give contradicting results, especially as far as the observability
of multiple gaps is concerned. Here we will show that, however, raw
data measured in different kinds of samples by different groups can
give surprisingly consistent results if they are treated in the same
way. Figure \ref{fig:Sm1111} (a) shows the raw conductance curves
measured by using the ``soft'' PCAR technique in a optimally doped
SmFeAsO$_{0.8}$F$_{0.2}$ polycrystal with $T_c=52$ K. We can notice
that: i) the low temperature curve shows maxima at about $\pm 5$ mV,
plus additional structures (here in the form of shoulders) at about
$\pm 15$ mV; ii) the normal-state conductance (thick line,
corresponding to $T=52.1$ K) is not flat but shows a residual,
asymmetric ``hump'' around zero bias, which is progressively
smoothed on increasing temperature. In \emph{all} the soft point
contacts we made in these polycrystals, this hump disappears --
leaving a flat, though still asymmetric conductance -- at about 140
-150 K (which is the temperature of magnetic ordering in the parent
compound) \cite{daghero09b,gonnelli09b}. The origin of this
``crossover'' is unclear. One could doubt that it might be related
to some residual magnetic order, but no trace of a secondary
magnetic phase with T$_N \approx 140$ K has been observed in our
samples at optimal doping (see e.g. \cite{Matusiak09}). A structural
transition has been claimed to occur between 175 K and 125 K along
the whole phase diagram of Sm-1111 \cite{martinelli11}, but we don't
have evidence of this effect in the samples studied here. A small
anomalous kink in the resistivity at similar temperatures, common to
1111 compounds (see e.g. \cite{kamihara10}), has been instead
observed, but its origin is not well understood yet. Transport
measurements \cite{chu10} and ARPES \cite{yi11} in detwinned
BaFe$_2$As$_2$ crystals showed that in-plane electronic anisotropy
occurs at temperatures above the structural transition, thus
indicating the presence of fluctuation effects \cite{chu10} which
have been interpreted as spin anisotropy in the paramagnetic phase
\cite{harriger11}. It might also be that similar effects are playing
a role in the 1111 systems and are possibly connected to the kink in
the resistivity and to the shape of the point-contact conductance.
However, detailed studies have to be performed as soon as bigger
crystals are available to clarify these points. Generally speaking,
we can say that the proximity of superconductivity to the AF SDW
state of the parent compound suggests the persistence, in the doped
compound, of spin fluctuations which could possibly give rise to the
observed behavior of the normal state at high temperatures. However,
since there are no clear predictions concerning the expected shape
of the PCAR conductance spectra and of their evolution with
temperature in the situations reported above, further experimental
and theoretical investigation should be performed in order to
clarify this issue. Then, it could also be possible to better
understand whether the different shape of the normal-state
conductance in La-1111 and Sm-1111 polycrystals is due to different
causes or it is only a different appearance of the same phenomenon.

Anyway, the anomalous shape of the normal-state conductance makes
again the normalization be a problem. Dividing by the curve at
$T_c^A$ is here a rougher approximation than in La-1111 because the
critical temperature is higher and one could reasonably suspect that
the normal state at $T_c^A$ can differ very much from that at low
temperature. The result of such a normalization is shown in panel
(c) for the curve at the lowest temperature. The normalized curve
still presents a strong right-left asymmetry, that some authors
remove by a symmetrization procedure. We chose instead to separately
fit the left and right side of the curves with the two-band 2D BTK
model, as shown by the lines superimposed to the symbols in panel
(c). The values of the small gap $\Delta_1$ extracted from the two
fits (5.7 meV) coincide, but the values of $\Delta_2$ are rather
different: 17.0 and 19.0 meV, respectively.

The fit of the symmetrized conductance obtained from the one
reported in figure \ref{fig:Sm1111}(c) gives $\Delta_1 = 5.7$ meV
and $\Delta_2 = 18$ meV. While the small gap remains the same the
large gap is actually an average of those obtained from the separate
fit of the right and left side of the unsymmetrized one. The
symmetrization is thus an acceptable practice when the asymmetry of
the conductance curve is not too large. This is what happened in
case of Co-doped Ba-122 (see \cite{tortello10}) where we symmetrized
the conductance spectra. Since in our results on Sm-1111 the curves
show a rather large asymmetry we consider more reliable, in that
specific situation, a separate fit.

The temperature dependence of the OPs (normalized as in figure
\ref{fig:La1111}(f)) obtained in this way is shown in panel (e) by
up (left side) and down (right side) triangles. Unlike in La-1111,
the gaps follow a BCS-like trend up to $T_c^A$, that coincides with
the bulk $T_c$ \cite{daghero09b}. The same panel also shows the gaps
obtained by fitting the negative-bias side of conductance curves
obtained in a different polycrystal, with a different doping content
($x=0.09$) and thus a smaller $T_c^A= 39$ K (open circles). Despite
these differences, once the temperature and energy scales are
normalized as in panel (e), the values of the gaps agree very well
with those obtained for $x=0.20$ \cite{daghero09b}. A few
measurements of soft PCAR spectroscopy were also performed in very
small Sm-1111 single crystals (about $200 \times 200
\,\mu\mathrm{m}^2$) with the current (mainly) injected along the
$ab$ planes \cite{karpinski09}. As usual, we fitted the
negative-bias and the positive-bias side separately to account for
the observed strong asymmetry of the spectra. At low temperature,
the gaps turned out to be $\Delta_1=6.45 \pm 0.25$ meV and
$\Delta_2=16.6 \pm 1.6$ meV, while $T_c^A=50.1$ K. The normalized
values of the gaps (actually, the values of the gap ratio $2\Delta
/k_B T_c^A$) are shown as a function of $T/T_c^A$ in figure
\ref{fig:Sm1111}(e) as squares and diamonds. The agreement of the
data with those obtained in polycrystals is remarkable. In general,
when all the data sets obtained in various polycrystals and crystals
are considered together, the values of the $2\Delta_1/k_B T_c^A$
agree very well with one another, varying only between 2.5 and 3
(i.e. within the thinner colored region in panel (e)). As for
$2\Delta_2/k_B T_c^A$, it is more scattered and varies at low
temperature between 7 and 9 (i.e. within the wider colored region).
This larger spread arises both from the asymmetry of the curves and
from the greater uncertainty in the determination of the large gap,
also related to the uncertainty introduced by the normalization or
to the possible presence of two large gaps of similar amplitude.

PCAR measurements in the needle-anvil configuration were performed
by Chen et al. in polycrystals with a reduced $T_c=43$ K with
respect to the optimal-doping one \cite{chen08b, chen09}. Most of
the low-temperature curves shown in \cite{chen08b} resemble rather
closely those measured by us; they present clear peaks related to a
small gap plus additional shoulders and structures. Unlike in our
measurements, the normal-state conductance above $T_c^A$ presents a
pseudogaplike upward curvature similar to that reported above for
La-1111. Prior to fitting their curves, the authors use a
normalization procedure that is equivalent to supposing that the
normal state is a straight line fitting the tails of the
unnormalized curve. The problem is that, the higher is the
temperature, the smaller is the voltage range where this
approximation holds (see Supplementary Information in
\cite{chen08b}). Moreover, this obviously implies disregarding
possible structures at energies much higher than that of the small
gap. Traces of the (presumable) large gap remain in the curves, but
become so small to be confused with minor features, possibly related
to the electron-boson coupling (see \ref{subsect:EBI} and
\ref{sect:EBIinFebased}). Instead, if the curves are divided by the
normal-state conductance, the features connected to the large gap
are enhanced with respect to other features at even higher energy.
The fit of the conductance curves performed by Chen et al. gives
indeed a single gap, which follows an almost perfect BCS temperature
dependence shown in figure \ref{fig:Sm1111}(e) as solid stars (note
that, because of the difference in critical temperature between
different data sets, the axes are normalized). In \cite{chen09} the
same authors notice that the energy of some extra features (which at
$T=4.5$ K occur at about 20, 40 and 55 meV) evolve in a BCS-like way
on increasing temperature so that, if properly normalized to its
low-$T$ values, all the curves are superimposed to the gap. On the
basis of what discussed in section \ref{subsect:EBI}, it is tempting
to interpret these features as the hallmarks of the strong
electron-boson coupling.

More recently, PCAR measurements in the needle-anvil configuration
have been performed in Sm-1111 films by Naidyuk et al.
\cite{naidyuk10}. Figure \ref{fig:Sm1111}(b) shows the temperature
dependence of one of their curves, which show very clear structures
related to the small gap and a pseudogaplike normal state. The
peculiar evolution of the tails of the curves as a function of
temperature seems to indicate a large excess conductance at high
energy, possibly due to electron-boson structures, as discussed in
section \ref{subsect:EBI}. Features connected to the
superconductivity are clearly present up to $T_c^A \simeq 30$ K.
This is the temperature where the resistance has dropped to about
80\% of the normal-state value, although the transition starts at
about 34 K. As in La-1111, the authors' normalization makes the
Andreev signal decay to 1 already at $\pm 12$ meV, thus flattening
all the structures at higher energy. This, again, allows a fit of
the small-gap features alone, and only to about 18 K; above this
temperature the normalization produces structures that cannot be
fitted by a BTK model. The gap values follow a BCS-like trend that,
if extrapolated, would give a critical temperature $T^* \simeq 21.6$
K. This, together with the low-temperature value $\Delta=3.35$ meV,
gives an almost perfect BCS ratio $2\Delta /k_B T^* = 3.6$. However,
the ``true'' critical temperature of the contact is $T_c^A \simeq
30$ K, that gives instead a gap ratio of only 2.6, \emph{which is
exactly the same we obtained for the small gap} (see figure
\ref{fig:Sm1111}(e)). Based on this similarity, we tried to
normalize the curves in \cite{naidyuk10} following our method, and
then divided all the curves by the presumable normal state at
$T_c^A$ (dashed curve in panel b). The result of this normalization
(and of a subsequent symmetrization) is shown by symbols in panel
(d). The curve has then been fitted to a single-band 2D BTK model
(dashed line) and to a 2-band 2D model (solid red line). It is clear
that the two-gap model allows a better fit of the small-gap
feature's width but also catches the position of the abrupt change
of curvature, which turns out to be related to the second large gap.
The values of the gaps given by this tentative fit are $\Delta_1 =
3.5$ meV and $\Delta_2= 11$ meV. Their values, properly referred to
the relevant $T_c^A$, are shown in panel (e) as green stars and
again turn out to be in very good agreement with ours. The failure
of the 2D BTK model to fit the tails of the curve in (d) is not too
worrying; the shape of these tails is indeed suggestive of strong
electron-boson structures in the conductance that could only taken
into account by inserting the energy-dependent order parameters into
the 2D BTK model (see section \ref{subsect:EBI}).

\subsection{``122'' compounds}\label{subsect:122PCAR}
In 122 compounds, the model for spin-fluctuation-mediated
superconductivity predicts a nodeless OP which, according to some
authors, can however evolve in a peculiar nodal symmetry, with
three-dimensional nodes on one hole-like Fermi surface, when the
latter acquires a more three-dimensional character \cite{suzuki11}.
In a recent paper, a relationship is claimed between $2\Delta_h/k_B
T_c$ (being $\Delta_h$ the gap on the hole-FS) and the occurrence of
nodes in the \emph{electron} FS \cite{maitichub11}. According to these
authors, both the hole-doped system (Ba,K)Fe$_2$As$_2$ and the
electron-doped system Ba(Fe,Co)$_2$As$_2$ present nodeless gaps
around optimal doping, while a nodal OP is very likely in the
isovalent-doped compound BaFe$_2$(As,P)$_2$ \cite{maitichub11,suzuki11}.

122 compounds are more suited to PCAR measurements than 1111
compounds, since they can be grown in the form of large single
crystals that can be cleaved more or less easily. This allows
directional measurements with the probe current injected either in
the basal plane or along the $c$ axis. Moreover, in some of these
compounds, ARPES measurements have provided valuable information on
the amplitude and symmetry of the gap on the various sheets of the
FS. This information generally confirms or supports the finding of
PCAR in these materials.

\subsubsection{(Ba,K)Fe$_2$As$_2$}
ARPES experiments in the hole-doped system (Ba,K)Fe$_2$As$_2$ at
$x=0.4$ (T$_c=37$ K) \cite{ding08,nakayama09} showed the presence
of four isotropic gaps: three of similar amplitude (between 11 and
13 meV) on the inner hole-like FS cylinder and on the two
electron-FS sheets around M, plus a small one ($\simeq 5.8$ meV)
on the outer hole FS sheet. Within the uncertainty of the
measurement, which is rather large, the gap amplitudes were shown
to be compatible with a symmetry $\Delta(k)=\Delta_0 \cos(k_x)
\cos(k_y)$ which is indeed the functional form of the $s\pm$
symmetry in the reciprocal space. The gaps seemed not to follow a
BCS-like curve as a function of temperature.

The first directional PCAR measurements in the same system were
carried out by Samuely et al. \cite{samuely09b,szabo09} who used
single crystals of Ba$_{0.55}$K$_{0.45}$Fe$_2$As$_2$ with
superconducting transition between $T_c^{on}=30$ K and
$T_c^{zero}=27$ K. By injecting the current along the $ab$ planes,
they obtained spectra with very clear two-gap structures and no
trace of zero-bias peaks or maxima, thus ruling out the possibility
of any symmetry of the OP involving a change of sign or zeros on a
single FS sheet. The conductance curves were systematically
normalized to the conductance curve at $T_c^A$, which sometimes
shows a humplike structure (similar to that observed by us in
Sm-1111 polycrystals) and sometimes a pseudogaplike, V-shaped
feature. For current injection along the $c$ axis, a similar
pseudogaplike feature and no Andreev signal were observed. The
zero-bias conductance minimum was shown to be progressively filled
on increasing temperature, up to the temperature (about 85 K in
these samples) where weak signs of the onset of a magnetic order are
found in resistivity and specific heat \cite{samuely09b}. The most
reasonable interpretation of these findings is that magnetic order
and superconductivity coexist and are probed sometimes together,
sometimes exclusively, by PCAR \cite{samuely09b}. This possibility
is confirmed by direct evidences of such a phase separation on a
scale of about 50 nm \cite{park09}, and also by successive PCAR
measurements in Ba$_{0.60}$K$_{0.40}$Fe$_2$As$_2$ crystals with
$T_c\simeq 37$ K carried out by Lu et al. \cite{lu10}. These authors
indeed observed a typical V-shaped conductance valley with a
universal functional form $G(V)= G(0) + c|V|^n$ being $n\simeq 2/3$
in a fraction of their $c$-axis contacts on this compound, but also
in the nonsuperconducting parent compound BaFe$_2$As$_2$, as well as
in $\mathrm{(Sr_{0.6}Na_{0.4})Fe_2 As_2}$ ($T_c=36$ K) and
$\mathrm{Sr(Fe_{0.9}Co_{0.1})_2 As_2}$ ($T_c \simeq 22$ K).

\begin{figure*}
\includegraphics[width=\textwidth]{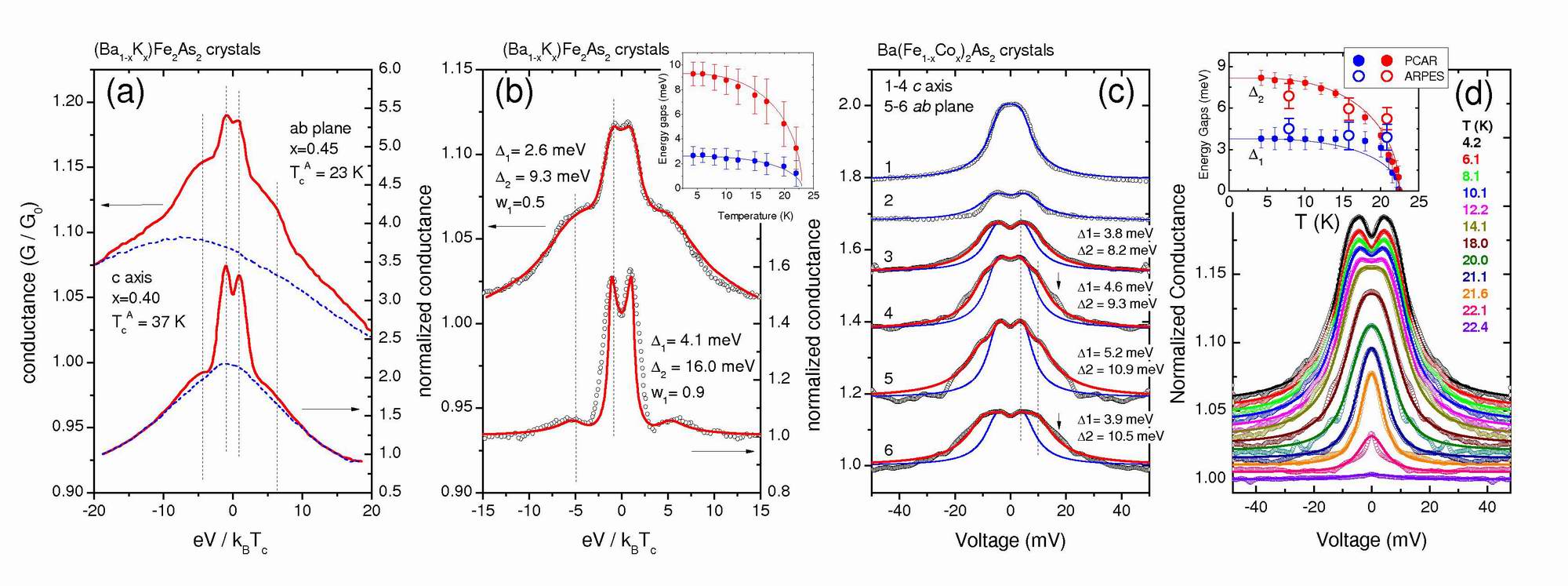}\\
\caption{(color online): (a) PCAR spectra obtained in
$\mathrm{Ba_{1-x}K_{x}Fe_2As_2}$ single crystals by injecting the
current along the $ab$ plane \cite{szabo09,samuely09b} and along
the $c$ axis \cite{lu10}. The curves are plotted as a function of
$eV/k_B T_c^A$ because the contacts had different $T_c^A$. Dashed
lines represent the normal-state conductance at $T_c^A$. (b) The
same curves after normalization (symbols) and the relevant fit
with a two-gap 2D BTK model (lines). The values of the gaps and of
the weight $w_1$ are reported in labels. Vertical lines indicate
the position of the gap structures. Inset: temperature dependence
of the gaps measured along the $ab$ plane
\cite{szabo09,samuely09b}. (c) Low temperature curves measured in
$\rm{Ba(Fe_{1-x}Co_{x})_2 As_2}$ single crystals, with current
along $c$ (curves 1-4) and $ab$ (curves 5 and 6). From
\cite{samuely09b} (curves 1-2) and \cite{tortello10} (curves 3-6).
Thin (thick) lines represent the 1-gap (2-gap) 2D BTK fit. Arrows
indicate bosonic structures. (d) Experimental conductance curves
of a $c$-axis contact as a function of temperature, with the
relevant 2-gap fit. The gap values are reported in the inset.
}\label{fig:Ba122}
\end{figure*}

Let us go back to the $ab$-plane measurements in
(Ba,K)Fe$_2$As$_2$ by Samuely et al. After the normalization, the
conductance curves were fitted to a two-band 2D BTK model with two
isotropic gaps. In different contacts, the small gap $\Delta_1$
was found to vary between 2 and 5 meV, and the large gap
$\Delta_2$ between 9 and 10 meV. The temperature dependence of the
gaps was found to agree very well with BCS-like curves. The
beautiful spectrum shown in figure \ref{fig:Ba122}(a) gives
$2\Delta_1 /k_B T_c^A \simeq 2.7$ and $2\Delta_2 /k_B T_c^A \simeq
9$, in very good agreement with the findings in La-1111 and
Sm-1111 shown in figures \ref{fig:La1111}(f) and
\ref{fig:Sm1111}(e). In some cases, small-amplitude curves with a
single broad maximum were obtained, that can be fitted by the same
model with very large values of the broadening parameters
$\Gamma_1$ and $\Gamma_2$. Let us note that these latter spectra
could be mistaken for evidences of nodal or anisotropic OPs if
there weren't clearer spectra ruling out this possibility. This
indicates how the cleanliness of the surface is essential for the
observation of clear gap structures (and thus a correct
interpretation of the data) as also pointed out by Lu et al.
\cite{lu10}, who performed $c$-axis PCAR measurements in
Ba$_{0.60}$K$_{0.40}$Fe$_2$As$_2$ crystals either cleaved or
uncleaved. In uncleaved crystals, they observed spectra with a
single, ill-defined maximum at zero bias and shallow structures at
about $\pm 15$ meV. Instead, in freshly cleaved samples two clear
conductance maxima were observed in the spectra. Once normalized
to the normal-state conductance (showing a humplike structure)
these spectra were fitted to a single-gap BTK model giving
$\Delta_1 = 3.0-4.0$ meV, corresponding again to $2\Delta_1 /k_B
T_c^A \simeq 2.6$ and thus in good agreement with Samuely's
measurements along the $ab$ plane. In addition to the clear
small-gap features, the raw spectra also presented smaller
structures that, in the normalized curve, show up as shoulders of
very small amplitude. These structures were interpreted by Lu et
al. as being ``dips'' due to the non-perfectly ballistic
conduction regime through the contact, in turn due to the very
small electron mean free path in these compounds \cite{lu10}. They
thus argued that $c$ axis PCAR measurements do not allow to detect
the large gap because it pertains to cylindrical FS sheets (the
inner-holelike at $\Gamma$ and the two electron-like at $M$),
while the small gap opens up on a more 3D Fermi surface sheet
\cite{wang09}.

In figure \ref{fig:Ba122} we compare two representative spectra
taken along the $ab$ plane \cite{samuely09b,szabo09} and along the
$c$ axis \cite{lu10}, by plotting them as a function of the
normalized energy $eV /k_B T_c$ (this is necessary because of the
different $T_c$'s of the samples). The similarity between the
spectra is striking; not only the small-gap features, but also the
additional structures now lie in the same position. This suggests
that the shallow structures at $\pm 15$ meV observed in the
$c$-axis spectrum might not be an artifact (as a matter of fact,
they always occur at the same voltage irrespective of the junction
resistance \cite{lu10}) but might instead be related to the large
gap. If this is the case, their small amplitude may be explained
by the much smaller ``weight'' of the relevant bands than in the
$ab$ plane (see section \ref{subsect:FS}). The normalized curves
are reported (again as a function of $eV /k_B T_c$) in figure
\ref{fig:Ba122}(b) as open symbols. The upper curve is compared to
the 2-band fit by Szab\'{o} et al. \cite{szabo09}, and the
relevant gap amplitude and weight are reported. The lower curve is
instead compared with a two-band 2D BTK fit that gives the same
small gap obtained by Lu et al. ($\Delta_1=4.1$ meV) but also a
large gap $\Delta_2 \simeq 16$ meV. Note that the weight of the
bands featuring the large gap is now only 0.1. Although this value
of $\Delta_2$ seems huge, the ratio $2\Delta_2/k_B T_c^A$ is the
same as for the $ab$-plane curve. If our interpretation is
correct, the available PCAR experiments in (Ba.K)Fe$_2$As$_2$ thus
concur to a picture of multiple nodeless gaps, in agreement with
the findings of ARPES. They also seem to indicate that these gaps
are isotropic (at least, this is what one would infer from the
directional invariance of the gap ratios $2\Delta_{1,2}/k_B
T_c^A$).

\subsubsection{Ba(Fe$_{1-x}$Co$_x$)$_2$As$_2$}
The first results of PCAR measurements in
Ba(Fe$_{0.93}$Co$_{0.07}$)$_2$As$_2$ single crystals with $T_c =
23$ K are reported by Samuely \emph{et al.} \cite{samuely09b}. The
$c$-axis spectra show clear Andreev signals, and in one case a
double-peak structure, but no evidence of multiple gaps. They were
thus fitted to a single-gap, $s$-wave BTK model giving $\Delta=
5-6$ meV which corresponds to $2\Delta/k_B T_c^A = 5.3 - 6.3$
since $T_c^A=22$ K. A large broadening parameter ($\Gamma\geq 0.5
\Delta$) was necessary to fit the curves. An example of these
curves is shown in figure \ref{fig:Ba122}(c) (curves 1 and 2). In
\cite{lu10}, also Lu \emph{et al.} reported on $c$-axis PCAR
measurements on Ba(Fe$_{0.9}$Co$_{0.1}$)$_2$As$_2$ crystals. They
observed V-shaped conductance curves with no Andreev signal when
the tip was gently put in contact with the cleaved sample surface,
and a zero-bias conductance enhancement when the pressure was
increased sufficiently. This feature is clearly related to
superconductivity since it disappears at $T_c$, but its shape
cannot be reconciled with an isotropic gap. One could think that
this feature is rather suggestive of a $d$-wave order parameter,
and try to fit it with a suitable generalized model.
%
%
However, the current is here injected along the $c$ axis. In this
configuration, no interference effects are expected for the $d$-wave
symmetry (with nodes on the hole FS), for the so-called ``nodal s''
symmetry (with nodes on the electron FS) \cite{kuroki09} and even
for the particular 3D nodal symmetry specifically predicted for 122
compounds \cite{suzuki11}. More exotic symmetries (if any) should be
imagined to make HLQ and ELQ experience a sign change of the pairing
potential in this configuration. Alternatively, one could imagine
the zero-bias maximum to simply arise from the anisotropy of the OP
but in this case, as shown in \ref{subsect:FS}, a closed 3D FS has
to be invoked, which is not predicted by bandstructure calculations.

We recently performed PCAR measurements with the ``soft''
technique in single crystals of Ba(Fe$_{0.9}$Co$_{0.1}$)$_2$As$_2$
(bulk $T_c^{on}=24.5$ K, $\delta T_c=1$ K). We  obtained various
series of spectra by injecting the current along the basal plane
or along the $c$ axis \cite{tortello10}. The normal-state
conductance at $T_c^A$ presented a shallow V-shape and
progressively flattened on increasing temperature. Figure
\ref{fig:Ba122}(c) reports some examples of normalized  (i.e.
divided by the conductance at $T_c^A$) and symmetrized curves for
$c$-axis (curves 3,4) and $ab$-plane contacts (curves 5,6). They
all feature maxima related to a small gap, plus additional
shoulders that can be related to a second, larger gap. The
position of these features does not depend on the direction of
current injection and this speaks in favor of isotropic gaps. In
this concern, it is worth mentioning that also the relative weight
of the two terms in the conductance (let's say, $w_1$ and
$(1-w_1)$) is the same irrespective of the direction of the
current. This seems to indicate that all the FS sheets have a
similar degree of three-dimensionality, as also shown by ARPES
\cite{vilmercati09} and x-ray Compton scattering \cite{utfeld10}.
Thick lines in figure \ref{fig:Ba122}(c) represent the two-gap 2D
BTK fit of the curves, while thin lines indicate the single-gap 2D
BTK fit. It is clear that the latter model is unapt to reproduce
the shape of the curves. The two-gap model works better in the gap
region but leaves out additional structures (indicated by arrows)
which are very clear only in curves with a large amplitude and
that can be related to the strong electron-boson coupling (see
sections \ref{subsect:EBI} and \ref{sect:EBIinFebased}).
Incidentally, note that in curve 5 these structures are probably
masked by small dips, that however do not affect the clear
double-gap structures. The absence of zero-bias peaks certainly
rules out, also in this case, the presence of nodes or line zeros
on the Fermi surface, and this is consistent with theoretical
predictions \cite{suzuki11}. Instead, these findings are not
incompatible with the presence of gap minima or even zeros (but
not line zeros) in some regions of the Fermi surface, as recently
proposed to explain the Raman scattering data in this compound
\cite{mazin10}.

The gaps obtained by PCAR can now be compared with the results of
ARPES measurements in the superconducting state. Terashima et al.
\cite{terashima09} measured a large isotropic gap $\Delta_2\simeq
7 \pm 1$ meV on the hole-like FS sheet around $\Gamma$ and a
smaller isotropic gap $\Delta_2\simeq 5 \pm 1$ meV on the
hybridized electron-like FS sheets around M. The values of the
gaps as a function of temperature are shown as open circles in the
inset to figure \ref{fig:Ba122}(d). As for the values of
$\Delta_1$, ARPES and PCAR agree rather well, while the value of
$\Delta_2$ is rather different, although the error bars overlap. A
possible explanation of this difference can be found in
\cite{tortello10}. Also note that the ARPES gaps show a weaker
temperature dependence and go to zero abruptly; the same result
has been found in $\mathrm{Ba_{1-x}K_{x}Fe_2As_2}$ \cite{ding08}.

\subsubsection{Ca(Fe$_{1-x}$Co$_x$)$_2$As$_2$}
\begin{figure}
\includegraphics[width=\columnwidth]{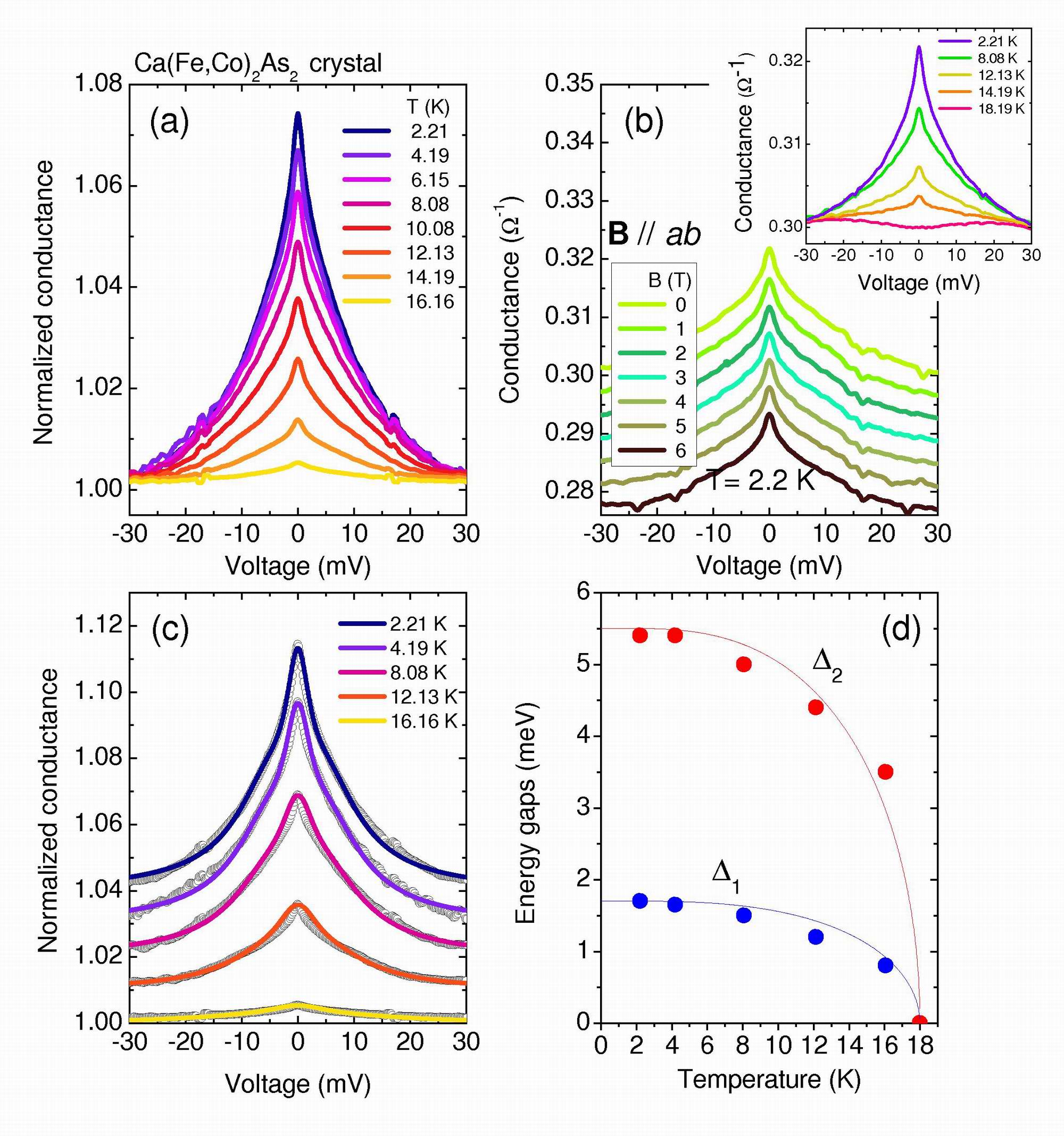}\\
\caption{(color online): (a) PCAR spectra obtained in
$\mathrm{Ca(Fe_{1-x}Co_{x})_2As_2}$ single crystals by injecting the
current along the $ab$ plane, normalized to their normal state at
$T_c^A = 18$ K. (b) Magnetic-field dependence of the same spectra,
at $T=2.2$ K. The inset shows some raw conductance curves at
different temperatures. (c) Some of the spectra of panel (a) with
the relevant two-band 2D BTK fit with a large isotropic gap
$\Delta_2$ and a small $d$-wave gap $\Delta_1$. (d) Temperature
dependence of the gaps as obtained from the fit of panel (c). Lines
are BCS-like curves.}\label{fig:Ca122}
\end{figure}

Very recently, we have performed PCAR measurements on
Ca(Fe$_{1-x}$Co$_x$)$_2$As$_2$ single crystals, with a bulk
critical temperature $T_c=18$ K as determined from DC
susceptibility measurements. The ``soft'' point contacts were made
on the freshly exposed side surface of a single crystal. Figure
\ref{fig:Ca122}(a) shows the temperature dependence, from 2.21 K
up to 16.16 K, of the conductance curves of a $ab$-plane point
contact ($R_N \simeq 3$ $\Omega$) normalized to the normal-state
conductance at $T_c^A=18$ K. The conductance curves are very
different from those observed in the other compounds presented so
far. First of all, they have very small amplitude; then, they
systematically present (in 100 \% of the cases) a zero-bias
enhancement of the conductance, similar to that occasionally
observed for example in $\mathrm{Ba_{1-x}K_{x}Fe_2As_2}$
\cite{lu10} and in Nd-1111 \cite{yates08a}; third, they have a
rather broad, almost triangular shape. These characteristics do
not change sensibly when a magnetic field up to 6 T is applied
parallel to the $ab$ plane, as shown in panel (b). This rules out
magnetic scattering and intrinsic Josephson junctions as the
origin of the peak and indicates that the upper critical field is
much higher than the maximum applied field. Furthermore, the
systematic occurrence of the ZBCP, the similarity between
different series of curves obtained in contacts with different
resistance and the absence of effects normally observed in
non-ballistic contacts, point towards an intrinsic origin of the
ZBCP.

As shown in section \ref{subsect:FS} a zero-bias peak or maximum
in \emph{ab}-plane contacts can be a sign of a nodal OP. Let's
assume that this is the case. This could be in agreement with the
predictions by Suzuki \emph{et al.} \cite{suzuki11} that, in 122,
nodes appear on the quasi-3D hole-like FS around the $\Gamma$
point (called $\alpha_1$ in \cite{suzuki11}) when the height of
the pnictogen on the Fe layer is small. In the similar compound
$\mathrm{Ba(Fe,Co)_2 As_2}$ \cite{sefat09} bandstructure
calculations show indeed that one of the holelike cylinders is
strongly warped and that its diameter grows considerably going
from $\Gamma$ to $Z$, which is the situation envisaged in
\cite{suzuki11}. However, in the absence of specific data about
the FS of Ca(Fe$_{1-x}$Co$_x$)$_2$As$_2$ and of detailed
information on the fine structural details, we must rely on the
PCAR spectra alone. In the absence of a specific way to express
the peculiar nodal symmetry predicted for 122 compounds, we just
tried to see whether a simple nodal symmetry (i.e. the $d$-wave
one) can account for the observed PCAR results. Note that quite
similar results can be however obtained by fitting with a fully
anisotropic $s$-wave OP as the one discussed in section
\ref{subsect:FS}. To reproduce the ZBCP a misorientation angle
$\alpha$ between the lobes of the $d$-wave gap and the normal to
the interface has to be included in the 2D BTK model (which is the
simplest approximation, justified by the roughness of the fit). We
were able to fit the conductance curves at any temperature up to
$T_c^A=18$ K by using a small $d$-wave gap and a larger $s$-wave
gap (presumably associated to the electron FS sheets). As usual,
we kept the barrier parameters $Z_1=0.5$ and $Z_2=0.285$ as well
as the weight $w_1=0.55$ and the angle $\alpha= \pi/9$ independent
of temperature. The fit of a subset of the curves in figure
\ref{fig:Ca122}(a) is shown in figure \ref{fig:Ca122}(c), and the
relevant gaps are shown as a function of temperature in panel (d).
Although the statistics is still to be completed, these results
may suggest the Co-doped Ca-122 system as a possible good
candidate for the study of nodal (or fully anisotropic)
superconductivity in 122 compounds.

\subsection{``22426'' compounds}\label{subsect:22426PCAR}
Evidences of nodal OP have also been reported for another class of
compounds, the highly layered ``22426'' family. Yates et al.
\cite{yates10} performed PCAR experiments, either with the
conventional ``needle-anvil'' technique or the ``soft'' technique
in $\mathrm{Sr_2ScFePO_3}$ polycrystals with $T_c=17$ K. This
material is particulary interesting, though less known than other
Fe-based compounds, for the study of the OP symmetry.
%
%
As a matter of fact, in 1111 \cite{kuroki09} and 122 \cite{suzuki11}
the height of the pnictogen (here P) above the Fe plane has been
indicated as the responsible for the appearance of nodes in the OP
(in 1111 compounds, this is accompanied by a decrease of $T_c$ which
does not necessarily occur in 122). In $\mathrm{Sr_2ScFePO_3}$ the
angle of the Fe-P-Fe bond ($118^{\circ}$) and the \emph{a}-axis
lattice parameter (4.016 {\AA}) mean that the height of the P above
the Fe plane is 1.20 {\AA}, which lies between that of LaFeAs(O,F)
and that of LaFePO (in which some evidences of nodal
superconductivity have been found).

\begin{figure}
\includegraphics[width=\columnwidth]{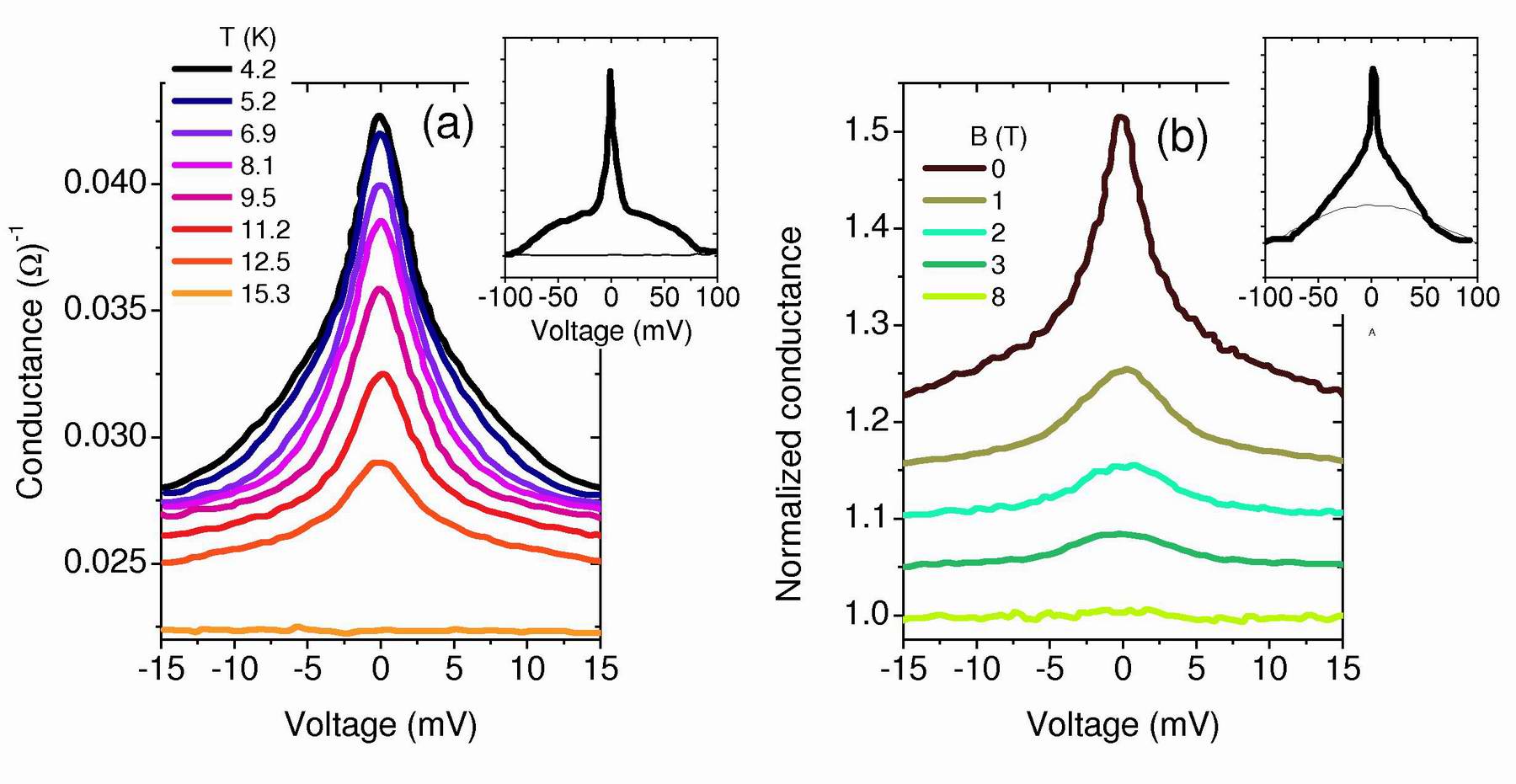}\\
\caption{(color online): (a) PCAR spectra obtained in a
$\mathrm{Sr_2ScFePO_3}$ polycrystal by Yates et al. \cite{yates10}
at different temperatures. The curves are not offset. Inset: the
low-temperature raw conductance (top) and the normal-state one at
$T_c^A$ (bottom). (b) Magnetic-field dependence (at 9 K) of the
conductance curves of a different point contact. All the curves
were normalized to their value at 20 meV. Inset: the raw
conductance in zero field (top) and at $H_{c2}$ (bottom).
}\label{fig:yates}
\end{figure}

The spectra obtained in \cite{yates10} systematically show a
zero-bias peak irrespective of the junction resistance (ranging from
30 to 100 $\Omega$) and of the PCAR technique used.
Fig.\ref{fig:yates}(a) shows the temperature dependence of one of
the spectra obtained with a tip pressed against the sample surface.
The curves are not offset; the apparent shift is due to a large
conductance excess that extends up to very high energies. The inset
shows indeed that it is only at $eV> 100$ meV that the tails of all
the curves are superimposed. It is also clearly seen that the normal
state conductance at $T_c^A$ is approximately flat. If this curve is
used to normalize the low-temperature conductances, the resulting
curves will look very similar to the raw ones, apart from the
vertical scale. It is clear that a fit of these curves in the whole
voltage range is impossible with any version of the BTK model with
energy-independent order parameters.

Figure \ref{fig:yates}(b) shows the magnetic-field dependence of the
conductance curves (unfortunately, of a different contact) at $T=9$
K, normalized at 20 meV \cite{yates10}. The ZBCP is progressively
depressed by the field, as well as the overall Andreev signal. A
relatively small field of 7 T is sufficient to completely suppress
superconductivity. $\mathrm{Sr_2ScFePO_3}$ is thus peculiar since
the normal state conductance at $T<T_c^A$ \emph{is} experimentally
accessible and can in principle be used to normalize the data. The
inset reports indeed the raw conductance curve at 9 K in zero field
(top) and at the upper critical field (bottom). The latter is
different from the normal state at $T_c^A$ shown in the inset to
panel (a) but unfortunately the two sets of curves refer to
different contacts and it is not possible to draw conclusions in
this regard. \\
A detailed analysis of the field-dependence \cite{yates10} shows
that the ZBCP is likely to have an intrinsic origin, apart from a
small contribution from intergrain weak links which is suppressed
already at 1 T.
However, the attempts to fit the conductance curves with a single
$d$-wave gap were not completely satisfactory due to the residual
excess conductance at high bias (that persists even if one uses the
normal-state conductance at $H_{c2}$ to normalize the curves). Even
a two-gap model fails unless abnormally high values of the large gap
are used to simulate the excess conductance. Only when choosing a
normalization that removes this excess, the authors obtained a
fairly good fit of the central peak that gave $\Delta=4.34 \pm 0.04$
meV, corresponding to a gap ratio $2\Delta/k_BT_c\simeq 6.7$.

\section{Electron-boson interaction in Fe-based
superconductors}\label{sect:EBIinFebased} As explained in detail in
section \ref{subsect:EBI}, in moderate- or strong-coupling
superconductors the PCAR conductance curves can show signatures of
the energy dependence of the superconducting gap. These structures
are more easily observable if the amplitude of the Andreev signal is
large; their signature in the second derivative of the $I-V$ curve
can be related to the electron-boson spectral function. In the
following, we will show some results obtained in Ba-122 single
crystals and Sm-1111 polycrystals that provide examples of such
strong-coupling effects and of their analysis. Especially for the
case of Ba-122 compounds, when compared to inelastic neutron
scattering measurements, these results strongly support a
spin-fluctuation-mediated origin of superconductivity.

\subsection{"122" compounds}\label{EBI_122}
\begin{figure}[h]
\begin{center}
\includegraphics[keepaspectratio, width=\columnwidth]{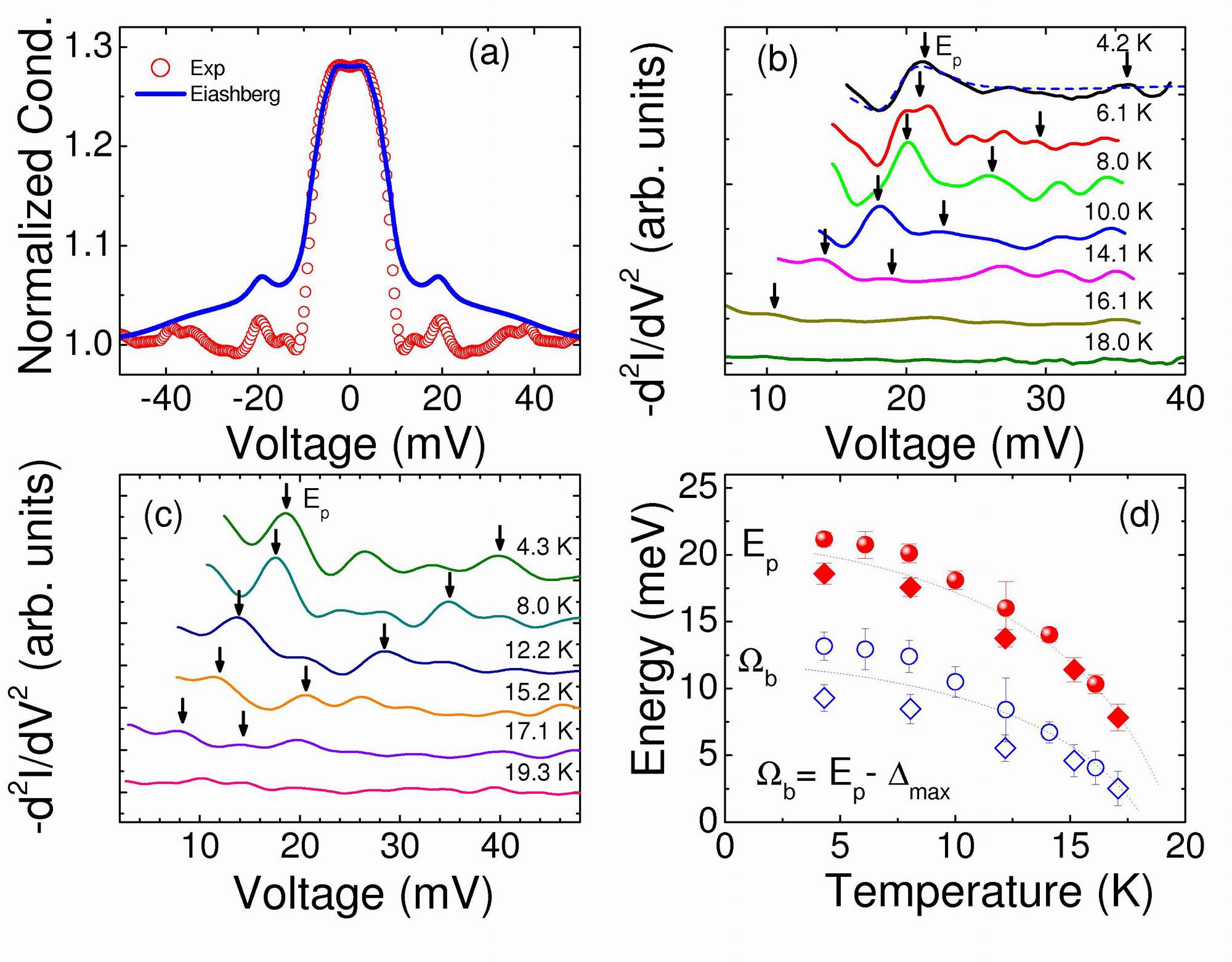}
\end{center}
\caption{(color online) (a): Normalized experimental conductance
curve (circles) obtained in a Ag/BaFe$_{1.8}$Co$_{0.2}$As$_{2}$
point-contact together with the theoretical one (line) obtained by
introducing in the BTK model the energy-dependent superconducting
gaps. The gaps as obtained by the Eliashberg calculations are
$\Delta_1=6.1$ meV, $\Delta_2=-3.8$ meV and $\Delta_3=-8.0$ meV
\cite{tortello10}. (b): Temperature dependence of the $-d^2I/dV^2$
curves obtained from the same contact of (a) showing the
displacement of the bosonic structures with increasing
temperature. Dashed line is obtained from the theoretical curve
shown in (a). (c): same as in (b) but for another contact. (d):
Temperature dependence of the energy peak, E$_p$ (full symbols)
extracted from (b) and (c) together with the corresponding boson
energy $\Omega_b(T)=E_p-\Delta_{max}$. Lines are guide to the
eye.} \label{Fig12}
\end{figure}

Figure \ref{Fig12}(a) shows an experimental conductance curve
(circles) obtained on a Co-doped Ba-122 single crystal close to
optimal doping. It can be clearly seen that the curve features
additional structures at energies higher than the large gap,
around 20 mV and, (less clearly) at about 40 mV. The solid line
represents a theoretical conductance curve obtained by introducing
in the two-band 2D BTK model the energy-dependent superconducting
gaps calculated within a three-band Eliashberg model (further
details can be found in \cite{tortello10}) by using a Lorentzian
electron-boson spectrum with characteristic boson energy
$\Omega_0\!=\! 12$ meV, very similar to the spin resonance energy
observed in neutron scattering measurements \cite{inosov10}. The
theory can reproduce the shape and position of the additional
structure at 20 mV while the one at higher energy is only faintly
visible in the theoretical conductance curve. Figure
\ref{Fig12}(b) shows the temperature dependence of the
sign-changed second derivatives (solid lines) of the I-V curves of
the same contact. As it can be expected, the electron-boson
interaction features are now seen more clearly. The dashed line is
the derivative of the theoretical curve shown in figure
\ref{Fig12}(a): it reproduces very well the 20 mV-structure while
the other one, though present, is hardly visible. This could be
due to the fact that the actual shape of the electron-SF spectral
function is more complex than the simple Lorentzian used here. By
looking at the temperature dependence of the $-d^2I/dV^2$ curves
(another example is reported in figure \ref{Fig12}(c)) it is
possible to notice that, as expected, both structures shift in
energy on increasing temperature. The position of the lower-energy
peak as a function of temperature, E$_p(T)$ is reported in figure
\ref{Fig12}(d) for the two cases (full symbols). The decrease of
E$_p$ is at least partially due to the closing of the
superconducting gap with increasing temperature. It is thus
essential to extract the temperature dependence of the
characteristic energy of the boson spectrum,
$\Omega_b(T)=E_p(T)-\Delta_{max}(T)$ (see section
\ref{subsect:EBI}). Figure \ref{Fig12}(d) shows that the values of
$\Omega_b$ (open symbols) decrease on increasing temperature, and
tend to zero when $T \rightarrow T_c$. This indicates that
$\Omega_b$ cannot be the energy of a phonon mode (in that case it
would not tend to zero!) and thus rules out a phononic origin of
this feature. Instead, the trend of $\Omega_b$ is very similar to
that of the spin-resonance energy peak reported in \cite{inosov10}
and thus strongly supports a spin-fluctuation-mediated pairing
mechanism in these compounds.

In this regard it is worth noticing that the existence of a
relationship between T$_c$ and the spin-resonance energy observed
by inelastic neutron scattering in Fe-based compounds does not
necessarily prove the spin-fluctuation pairing. As J. Paglione and
R.L. Greene point out, since such a relationship appears
exclusively in nearly magnetic unconventional superconductors, it
could be considered an independent remnant of nearby magnetic
order rather than a strong signature of magnetically mediated
pairing \cite{paglione10}.

What PCAR spectroscopy provides is something more than that. Indeed:
i) it detects features related to the \emph{interaction of electrons
with elementary excitations} in the material (and not only to the
excitation itself); ii) it shows that these excitations enter the
imaginary part of the energy-dependent order parameter; iii) it
allows extracting the characteristic energy of the
\emph{interaction} spectrum $\alpha^2F(\Omega)$, which turns out to
coincide with that associated to spin fluctuations and observed by
neutron scattering experiments.

\subsection{"1111" compounds}

\begin{figure}[h]
\begin{center}
\includegraphics[keepaspectratio, width=0.7\columnwidth]{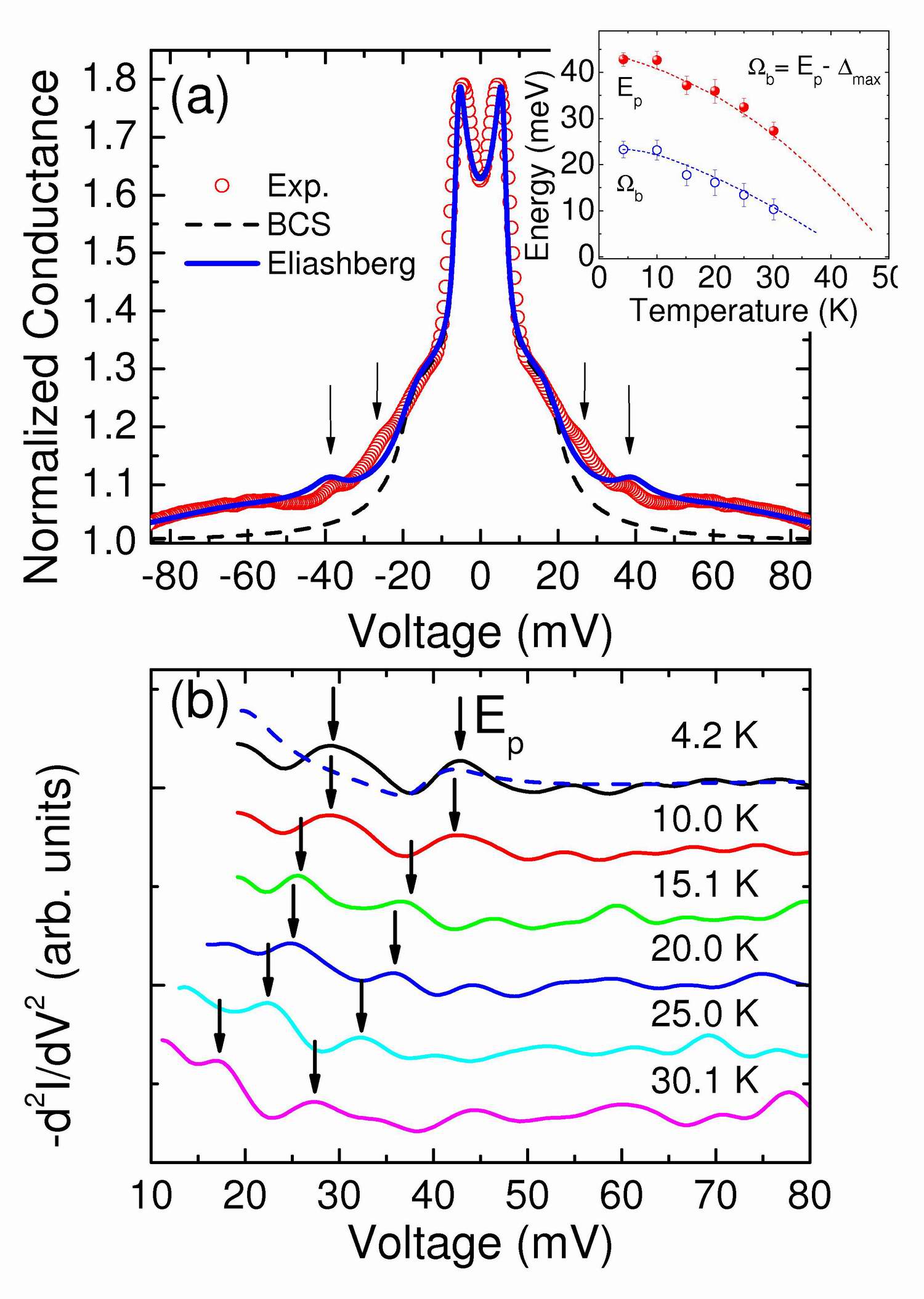}
\end{center}
\caption{(color online) (a): Normalized experimental conductance
curve (circles) obtained in a Ag/SmFeAsO$_{0.8}$F$_{0.2}$
point-contact. Dashed line is a BTK fit to the experiment,
obtained by using the constant BCS values for the gaps. The
parameters of the fit are $\Delta_1=6.0$ meV, $\Delta_2=19.5$ meV,
$\Gamma_1=0.1$ meV, $\Gamma_2=2.5$ meV, Z$_1$=0.2, Z$_2$=0.1 and
w$_1$=0.6. Solid line is a theoretical curve obtained by
introducing in the BTK model the energy-dependent gap functions
calculated within the three-band Eliashberg theory whose values
are $\Delta_1=6.03$ meV, $\Delta_=17.23$ meV and $\Delta_3=-19.56$
meV. (b): Temperature dependence of the $-d^2I/dV^2$ curves
obtained from the same contact of (a) showing the displacement of
the bosonic structures with increasing temperature. Dashed line is
obtained from the theoretical curve shown in (a). Inset:
Temperature dependence of the energy peak, E$_p$ (full symbols)
extracted from (b) together with the corresponding boson energy
$\Omega_b(T)=E_p(T)-\Delta_{max}(T)$. Lines are guide to the eye.}
\label{Fig13}
\end{figure}

Figure \ref{Fig13}(a) shows a normalized experimental conductance
curve obtained by us on a optimally doped Sm-1111 polycrystal
(circles). It looks very similar to the spectrum by Naidyuk \emph{et
al} \cite{naidyuk10} shown in figure \ref{fig:Sm1111}(d) and also to
some PCAR spectra in underdoped polycrystals by Chen et al.
\cite{chen09}. The amplitude of the Andreev signal in this contact
is exceptionally high (about 80\%) and, besides clear two-gap
features (peaks and shoulders), additional structures or small kinks
can be seen around 27 mV and 40 mV. The dashed line is a BTK fit to
the experiment by using the two-band 2D BTK model with BCS gap
values (i.e. independent of energy). As in figure
\ref{fig:Sm1111}(d), the fit reproduces very well the experiment in
the central part of the curve (and allows obtaining reliable values
of the gaps) but fails at higher energies. The solid line is instead
the result of inserting in the same BTK model the energy-dependent
OPs obtained by solving a three band Eliashberg model
\cite{ummarino10b} in which, as usual, the electron-boson spectral
function is modeled by a Lorentzian curve and the calculated gaps
are $\Delta_1=17.23$ meV, $\Delta_=6.03$ meV and $\Delta_3=-19.56$
meV.

Since to the best of our knowledge no spin-resonance energy value
is available for this compound, the characteristic energy has been
chosen according to \cite{paglione10} by extrapolating the
relationship $\Omega_0 \sim 4.65 k_B T_c \sim 20$ meV. Although
the theoretical curve shows no structures at 27 mV (which by the
way corresponds approximately to $\Delta_{min}+\Omega_0$), the
feature at 40 mV is remarkably well reproduced as can be observed
in figure \ref{Fig13}(a) and \ref{Fig13}(b) (dashed line).
Similarly to the case of Co-doped Ba-122, only the structure
present at approximately $\Omega_0 + \Delta_{max}$ is reproduced.
This again indicates that the model has to be investigated further
or that additional features of the spectral function are playing
an important role. As expected, both structures shift in energy on
increasing temperature, partly because the amplitude of the
superconducting gaps is also decreasing. The inset to figure
\ref{Fig13}(a) reports the position of the energy peak in the
second derivative, E$_p$ (full symbols) and the values of
$\Omega_b=E_p-\Delta_{max}$ (open symbols) as a function of
temperature. Similarly to the case of Co-doped Ba-122 reported in
the previous section, $\Omega_b$ decreases in temperature
indicating that the observed structure does not have a phononic
origin and might instead be related to spin fluctuations. It is
interesting to note that similar structures at about 20, 40 and 55
meV were directly observed in the PCAR spectra by Chen \emph{et
al.}, who also observed their shift in energy, similar to that of
the gap \cite{chen09}, but did not give them any special physical
meaning.

\section{The puzzle of $2\Delta/k_B T_c$ ratios}\label{sect:puzzle}
\begin{figure}[h]
\begin{center}
\includegraphics[keepaspectratio, width=\columnwidth]{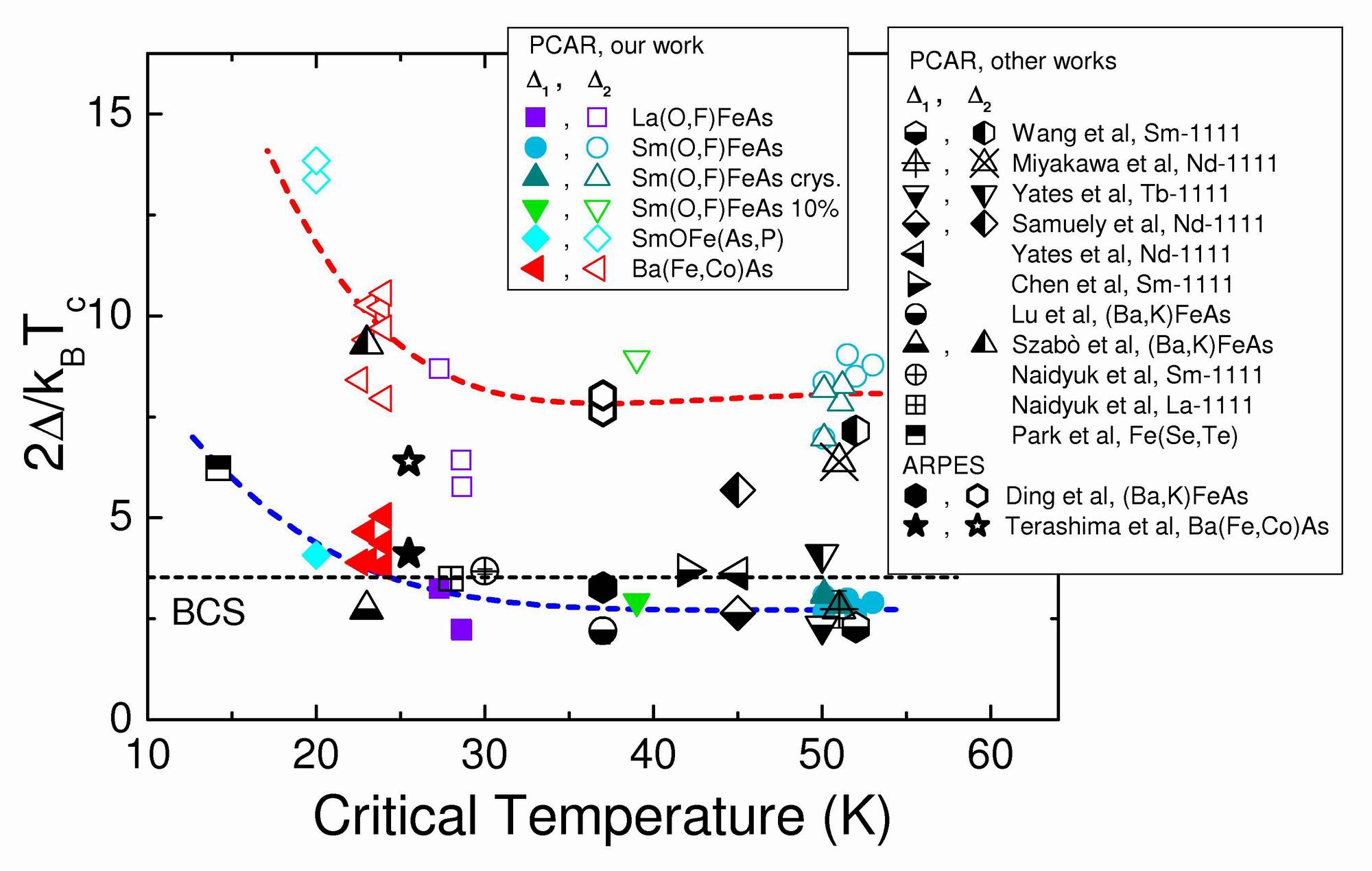}
\end{center}
\caption{(color online): $2 \Delta_1 / k_B T_c$ and $2 \Delta_2 /
k_B T_c$ values vs T$_c$ as determined by PCAR spectroscopy
measurements performed by our group as well as several other groups
in case of nodeless gaps. Horizontal dashed line represents the
s-wave weak coupling limit for the BCS theory. Blue and red dashed
lines are only guides to the eye. PCAR data for SmOFe(As,P) are
taken from \cite{zhigadlo11}, other results published in literature
are from
\cite{miyakawa10,yates09,samuely09a,yates08a,chen08b,lu10,szabo09,
naidyuk10,yates10,park10}. The data labelled ``Wang et al.''
indicate the two-band s-wave fit of the spectra from ref.
\cite{wang08}. ARPES data are from \cite{ding08,terashima09}. Data
from \cite{naidyuk10} are plotted vs $T_{c}^{A}$. } \label{Fig14}
\end{figure}

Figure \ref{Fig14} reports a collection of the values of the gap
ratios 2$\Delta_i/k_B T_c$ (where $\Delta_i$ is the $i$-th band gap
measured at low temperature) given by different PCAR measurements
performed by us and other groups in various Fe-based systems. Only
results referring to nodeless gaps have been reported essentially i)
because they represent the great majority the cases studied up to
now and ii) to allow an homogeneous comparison with the standard
$s$-wave BCS ratio. The gap ratios are reported as a function of the
$T_c$ of the sample or, when available, of the $T_c^A$ of the
contact. Two representative ARPES data are also shown for
comparison. Let us start by analyzing the trend of $2\Delta_1/k_B
T_c$, where $\Delta_1$ is, as usual, the small gap. In samples with
high $T_c$, this ratio is close to (but generally smaller than) the
BCS value for the $s$-wave weak-coupling regime. In most compounds,
it is around 2.5. It then remains almost constant when the $T_c$
decreases down to approximately 30 K; then, it starts to increase
continuously with further decreasing $T_c$.

A qualitatively similar trend is also observed for $2\Delta_2/k_B
T_c$, despite a larger spread of values which is not surprising
since the larger gap is worst resolved than the small one, its
absolute value could be affected by the choice of the normalization
and moreover, as pointed out in \ref{sect:EBIinFebased}, could also
be the average of two large gaps of similar value. We cannot,
however, exclude that the spread is also due to specific features of
each system such as nesting, vicinity to magnetism etc. but also,
and most probably, quality of the sample, homogeneity and disorder.

Although more data points are desiderable in the low-T$_c$ region of
the graph, figure \ref{Fig14} indicates that the increase of the
$2\Delta_i/k_B T_c$ ratio below a certain T$_c$ (on the order of 30
K) appears to be a rather common trend for both gaps. But, how can
this trend be interpreted?

In order to make a reliable statement in this regard, the complexity
of these systems has to be carefully taken into account. Because of
the multiband nature of these compounds and their dominant interband
pairing, a proper interpretation of the data would require a
quantitative analysis (for example within a multiband Eliashberg
model) taking into account the electronic structure of each
compound. However, as a first approximation, we have done numerical
simulations within the three-band s-$\pm$ Eliashberg theory (with
feedback effect \cite{ummarino10b}, i.e. by including in a
self-consistent way the temperature dependence of the spin
fluctuation spectral function) to see in what conditions this trend
can be reproduced. To simplify the problem, we fixed some parameters
to the values that allowed us to reproduce the gaps in Sm-1111
compound with T$_c$ = 52 K \cite{ummarino10b}. These parameters are
$\lambda_{23}/\lambda_{13}$ (where $\lambda_{ij}$ is the coupling
constant between bands $i$ and $j$), $n_{13}$ and $n_{23}$
($n_{ij}=N_i/N_j$ where $N_i$ is the density of states of band $i$).
According to \cite{paglione10}, we assumed $\Omega_0$ to vary
linearly with $T_c$, i.e. $\Omega_0 \simeq 4.65 k_B T_c$. For values
of the critical temperature down to 30 K, the experimental values of
$2\Delta_i / k_B T_c$ can be reproduced by keeping the total
coupling constant, $\lambda_{tot}$, practically constant to a value
of approximately 2.4. Below 30 K, to follow the experimental trend
it is necessary to i) increase $\lambda_{tot}$ and ii) modify the
temperature dependence of the peak energy. In standard situations,
the latter evolves with temperature according to
$\Omega_0(T)=\Omega_0\tanh[1.76 \sqrt{(T_c^*/T-1)}]$
\cite{ummarino10b} (i.e. in a BCS-like way)  where $T_c^*$ is the
critical temperature calculated \emph{without} the correction due to
the feedback effect ($T_c^* > T_c$). Here, we assume instead
$\Omega_0(T)=\Omega_0\tanh[1.76 k \sqrt{(T_c^*/T-1)}]$ where $k=1$
for $T_c \geq 30$ K and decreases for $T_c< 30$ K. Roughly speaking,
this is equivalent to a reduction of the representative frequency
$\Omega_0$ at any $T$ between 0 and $T_c^*$, with respect to the
above BCS-like curve. As a result, a larger $\lambda_{tot}$ is
necessary to obtain the correct $T_c$; but this also gives larger
gaps. This result, if confirmed by using detailed electronic
calculations for each compound, might indicate that below T$_c$=30 K
the coupling increases and would also have interesting implications
on the temperature dependence of the spin fluctuation spectral
function that, as shown in figure \ref{Fig12}(d) and in the inset of
figure \ref{Fig13}, can be determined from high-quality PCAR
spectroscopy experiments.

\section{Summary and conclusions}

In the present Report we have presented an overview of the (partly)
solved and of the (still) debated points about superconductivity in
Fe-based superconductors that have been (or can be) addressed by
means of point-contact spectroscopy. To do so, we have also
introduced generalizations of the usual models for the fit of the
PCAR spectra, and presented some new results. Here, we would like to
simply summarize the main conclusions of this long discussion, and
outline the possible future challenges for PCAR spectroscopy in the
research about Fe-based superconductors.

In Section \ref{sect:Models} we have examined and discussed in
detail the profound and often counterintuitive connections among the
FS topology of multiband superconductors, the symmetry of the OPs
present on the FS sheets, the directionality of the PCAR process,
the strength and the spectrum of electron-boson interaction, and the
resulting PCAR normalized conductance curves. Taking into account
these connections greatly complicates the description of the AR
process, but is unavoidable when dealing with Fe-based
superconductors. The extension of the BTK model to its most general
form, that we call ``full 3D BTK'' model, (sections \ref{subsect:FS}
and \ref{subsect:EBI}) allows directly comparing the PCAR
experimental results with the predictions of the theory for a
\emph{specific shape of the FS}, a \emph{specific symmetry of the
OP} in every band and a \emph{specific spectrum of electron-boson
interaction} (via the preliminary solution of the Eliashberg
equations). The comparison of the results to those of the simplified
BTK models provides some indications and warnings for the analysis
of PCAR data in Fe-based compounds that can be summarized as
follows:

1) The analysis of the \emph{ab}-plane PCAR spectra on Fe-based
superconductors requires at least the 2D version of the BTK model,
because of the quasi-2D shape of their FS and the possible presence
of non-isotropic OPs.  Even in the case of pure \emph{s}-wave
symmetry of all the OPs, this model gives results slightly different
from the full 3D one (in particular, an apparent enhancement of the
$Z$ values is observed, due to FS-shape-related geometric effects);

2) In principle, \emph{c}-axis PCAR spectra in pnictides should be
analyzed only by means of the 3D BTK model (simplified version of
\cite{yamashiro97} or full version of equation 11) because of the
non-spherical shape of the FSs. However if the \emph{ab}-plane
contacts on the same material don't show any sign of line nodes or
zeros in the OP, and the precise determination of the $Z$ values is
not vital, the 2D BTK model can return correct information about the
\emph{isotropic} gap values and their temperature dependency. In
this last case, as shown for the large gap in figure \ref{fig:4}(b),
the $Z$-enhancement effect is remarkable especially when the FS is
only slightly warped, but the gap-related peaks remain at the same
energy positions;

3) The application of the full 3D BTK model to the case of
\emph{fully anisotropic s} OPs (with line zeros) or \emph{d}-wave
OPs (with line nodes) that open on warped cylindrical surfaces leads
to interesting conclusions. As expected, the normalized conductance
of \emph{ab}-plane contacts in the AR regime ($Z<0.2-0.4$) shows
characteristic zero-bias maxima or peaks, respectively.
Unfortunately, in the presence of a small lifetime broadening the
smoothed spectra become very similar, preventing the possibility to
discriminate between the two symmetries. In the same conditions,
\emph{c}-axis contacts unexpectedly show no signs of zero-bias peaks
or maxima even for $Z=0$ because of the geometric $Z$-enhancement
effect, as shown in figure \ref{fig:4}. This prediction could be
verified by means of \emph{directional} PCAR experiments on
high-quality single crystals. In other words, the observation of
zero-bias peaks or maxima in \emph{c}-axis contacts can be
reconciled with the previous two OP symmetries only if the
corresponding FS sheet is 3D, thus giving to directional PCAR
spectroscopy the capability to attain indirect information on the
shape of the FS;

4) The introduction of energy-dependent gap functions $\Delta_i(E)$
(determined by the proper solution of multiband Eliashberg
equations) into the BTK expressions leads to the conclusion that
structures due to the electron-boson interaction (EBI) are certainly
visible in the AR conductance at energies higher than the large gap
(and for coupling constants compatible with the $T_c$ and the gap
values) \emph{provided that} the spectral function of this
interaction is sufficiently peaked (e.g. has a Lorentzian shape).
The use of the spectral function typical of antiferromagnetic spin
fluctuations (AFSF) results in curves with no EBI structures unless
unphysical, very low cutoff energies are chosen. Serious questions
thus arise on the true shape of the bosonic spectrum since (as shown
in two particular cases in section \ref{sect:EBIinFebased}) EBI
structures have actually been observed in the PCAR data of various
compounds.

As for the OP symmetry in Fe-based superconductors, the analysis and
comparison of various PCAR results reported in section
\ref{sect:PCAR} show that, despite the rather common opinion that
PCAR results are contradictory, the most recent and reliable
experiments are in fairly good agreement with one another. In
particular:

1) When data from various groups are available for a given compound,
they generally give similar results if analyzed in a homogeneous
way. The apparent conflicts are mostly due to a different
normalization of the raw data or to a different interpretation of
the common structures present in the conductance.

2) In optimally doped and slightly underdoped 1111 compounds, if the
PCAR curves are normalized to the relevant normal-state spectrum and
the high-energy structures are not artificially reduced, all the
data point to the presence of (at least) two isotropic gaps (with no
nodes and no line zeros) with ratios $2\Delta_i/k_BT_c$ lower and
sensibly higher than the BCS value, respectively.

3) Similar observations apply to the 122 compounds where data of
various groups concur in supporting a multi-gap $s\pm$ picture
(though at most two gaps can be resolved by PCAR spectroscopy).
Since in this case most of the experiments are made on single
crystals, a dependency of the ``weight'' of every band on the
direction of current injection is also observed, in agreement with
the predictions of subsection \ref{subsect:FS}.

4) In compounds of the 22426 family and in Co-doped Ca-122 (as shown
here for the first time) PCAR spectroscopy gives some evidence of a
\emph{d}-wave or a \emph{fully anisotropic s}-wave OP. In these
cases, experiments along different crystallographic directions can
be used as a ``test'' of the predictions of \ref{subsect:FS} or of
the presence of a 3D FS sheet.

In section \ref{sect:EBIinFebased} we have given evidence of  EBI
structures in the PCAR spectra of Co-doped Ba-122 and Sm-1111. By
properly combining the direct solution of the three-band $s\pm$
Eliashberg equations and the BTK models we were able to reproduce
these structures in amplitude and position, thus obtaining important
information on the characteristic energy of the bosonic spectrum and
its temperature dependency. This energy turns out to coincide with
the characteristic energy of the spectrum of spin fluctuations
measured by inelastic neutron scattering thus giving a \emph{strong
support} to the SF origin of the coupling in these compounds. In
addition these EBI-related structures (not only small peaks or
shoulders at the proper energy but also a general increase of the
conductance at high bias) can nicely explain some frequently
observed anomalies of PCAR spectra that cannot be reproduced by the
BTK models using constant gaps. This is another example that too
simplified models cannot explain the rather complex physics of AR in
Fe-based superconductors.

Finally, in section \ref{sect:puzzle} we have shown -- by comparing
the results of various experiments of different groups -- that at
the decrease of the $T_c$ of the sample the ratios
$2\Delta_i/k_BT_c$ for the small and the large gap have an anomalous
behavior. Instead of merging at a value close to BCS one at the
lowering of $T_c$, as naively expected in the framework of a
multiband model dominated by interband couplings, they both increase
at $T<30$ K reaching values greater than 5 and 10, respectively. A
preliminary analysis of this ``strange'' behavior in the framework
of the Eliashberg theory including feedback effect
\cite{ummarino10b} shows that it can be explained only supposing
that below $\sim 30$ K the temperature dependency of the
characteristic boson energy is no more BCS-like but becomes more
linear and, at the same time, the total coupling strength increases.
Both these conclusions could be further investigated by directional
PCAR experiments in low-$T_c$ high-quality single crystals.

In conclusion, we have shown that PCAR spectroscopy can provide
information on some debated properties of the Fe-based
superconductors if the PCAR spectra are analyzed within suitably
generalized models. In particular, a tighter relationship between
the electronic properties (and thus the geometry of the Fermi
surface) and the PCAR results can be established. This could help
either verifying the plausibility of calculated Fermi surfaces, or
providing some experimental hints about what the FS should look like
to explain the measured spectra. A systematic PCAR study of some
selected systems as a function of doping, coupled to bandstructure
and FS calculations as well as with the determination of fine
lattice details, could help testing the proposed connection between
these aspects and the symmetry of the order parameter, and to verify
to what extent the scaling law of the gap ratios $2 \Delta_i/k_B
T_c$ as a function of $T_c$ is universal in these compounds (and, if
so, what are its physical reasons and implications). As for the
origin of the coupling mechanism, important hints can be obtained by
extracting the typical boson energy from PCAR spectra and comparing
it with the energy of spin resonance as measured, for instance, by
neutron scattering experiment. In general, thus, the future
developments of the research about Fe-based compounds  pose a
challenge to PCAR spectroscopy, and might strongly benefit from the
results of this experimental technique.
\section*{Acknowledgments}
We would like to thank  Lilia Boeri, F. Dolcini, Laura H. Greene, S.
Massidda, I. I. Mazin, P. Raychaudhuri  for enlightening discussions
and useful suggestions. A particular acknowledgment to V. A.
Stepanov for his fundamental contribution to our research in all
these years. Many thanks to all those who grew and characterized the
samples we used for our measurements: J. Karpinski, N. D. Zhigadlo,
Z. Bukowski, J. Jiang, J. D. Weiss, E. E. Hellstrom, J.S. Kim, M.
Putti, and A. Palenzona. Finally, many thanks to R.K. Kremer and P.
Reuvekamp and the Max Planck Institute f\"{u}r
Festk\"{o}rperforschung (Stuttgart) where some of the measurements
reported here were carried out.

\end{document}